\documentclass[11pt]{article}
\usepackage[centertags]{amsmath}
\usepackage{amsfonts}
\usepackage{amssymb}
\usepackage{amsthm}
\usepackage{newlfont}
\usepackage{epsfig}
\usepackage{amscd}

\newcommand{\N}{\vec{N}}

\newcommand{\bpsi}{{\boldsymbol \psi}}

\newcommand{\RR}{{\mathbb R}}
\newcommand{\ZZ}{{\mathbb Z}}

\newcommand{\CC}{{\mathbb C}}

\newcommand{\bx}{{\boldsymbol x}}
\newcommand{\bv}{{\boldsymbol v}}
\newcommand{\bF}{{\boldsymbol F}}
\newcommand{\res}{\mathop{\mathrm{res}}}

\def\eqref#1{(\ref{#1})}
\theoremstyle{plain}
\newtheorem{Th}{Theorem}
\newtheorem{Cor}[Th]{Corollary}
\newtheorem{Lem}[Th]{Lemma}
\newtheorem{Prop}[Th]{Proposition}
\theoremstyle{definition}

\theoremstyle{remark}
\newtheorem*{Rem}{Remark}
\begin{document}
\title{
Integrable lattices and their sub-lattices: \\
from the discrete Moutard (discrete Cauchy-Riemann) \\ 4-point equation
to the self-adjoint 5-point scheme.
}
\author{
A. Doliwa\thanks{Uniwersytet Warmi\'{n}sko-Mazurski w Olsztynie,
Wydzia{\l} Matematyki i Informatyki,
ul.~\.{Z}o{\l}nierska 14 A, 10-561 Olsztyn, Poland,
e-mail: {\tt doliwa@matman.uwm.edu.pl}}~,
P. Grinevich\thanks{Landau Institute for Theoretical Physics, Moscow,
Russia,
e-mail: {\tt pgg@landau.ac.ru}}~,
M. Nieszporski\thanks{
Katedra Metod Matematycznych Fizyki,
Uniwersytet Warszawski
ul. Ho\.za 74, 00-682 Warszawa, Poland,
 e-mail: {\tt maciejun@fuw.edu.pl}},~and
P.M. Santini\thanks{Dipartimento di Fisica, Universit\`a di Roma
``La Sapienza'' and
Istituto Nazionale di Fisica Nucleare, Sezione di Roma,
Piazz.le Aldo Moro 2, I--00185 Roma, Italy,
e-mail: {\tt paolo.santini@roma1.infn.it}}
}
\date{}
\maketitle

\begin{abstract}
We introduce the sub-lattice approach, a procedure to generate, from a
given integrable lattice, a sub-lattice which inherits
its integrability features. We consider, as
illustrative example of this approach, the discrete Moutard 4-point equation
and its sub-lattice, the self-adjoint 5-point scheme on the star of
the square lattice, which are relevant in the theory of the integrable Discrete
Geometries and in the theory
of discrete holomorphic and harmonic functions (in this last context,
the discrete Moutard equation is called discrete Cauchy-Riemann equation). We use the sub-lattice
point of view to derive, from the Darboux
transformations and superposition formulas of the discrete Moutard equation,
the Darboux transformations and superposition formulas of the
self-adjoint 5-point scheme. We also construct, from algebro-geometric
solutions of the discrete Moutard equation, algebro-geometric
solutions of the self-adjoint 5-point scheme. We finally use these
solutions to construct explicit examples of discrete holomorphic and harmonic functions,
as well as examples of quadrilateral surfaces in $\RR^3$.

\end{abstract}


\section{Introduction}
One of the most important methods to generate integrable equations is
to start with a fairly general integrable system and
apply to it systematically symmetry reductions. This approach was,
for instance, applied to the multicompontent Kadomtsev--Petviashvili
hierarchy
\cite{KvL} and to the self-dual Yang-Mills system \cite{Ward}. On the level
of discrete equations, it was systematically used, for instance, to generate
various integrable reductions of the multidimensional quadrilateral lattice
equations (see, for example, \cite{DS}).

In this paper we propose a
different, but equally relevant, procedure to generate integrable lattices.
This procedure consists in constructing,
from a given integrable lattice, a sub-lattice which inherits the
integrability features of the original lattice.
We consider, as
illustrative example of this approach, the 4-point difference equation
\begin{equation}
\label{4p}
\phi_{m+1,n+1}-\phi_{m,n}=g_{m,n}(\phi_{m+1,n}-\phi_{m,n+1}),
\end{equation}
where $g_{m,n}$ and $\phi_{m,n}$ are functions: $\ZZ^2 \to \CC$, and
its sub-lattice, the self-adjoint 5-point scheme:
\begin{equation} \label{5p}
{a}_{\mu,\nu}\Psi_{\mu+1,\nu}+
{a}_{\mu-1,\nu}\Psi_{\mu-1,\nu}+
{b}_{\mu,\nu}\Psi_{\mu,\nu+1}+
{b}_{\mu,\nu-1}\Psi_{\mu,\nu-1}={c}_{\mu,\nu}
\Psi_{\mu,\nu}
\end{equation}
on the star of the square lattice,
where ${a}_{\mu,\nu}$, ${b}_{\mu,\nu}$,
${c}_{\mu,\nu}$ and $\Psi_{\mu,\nu}$
are functions: $\ZZ^2 \to \CC$, and where
the standard notation
$f_{m,n}=f(m,n)$ is often used throughout the paper.

We exploit the above fact deriving several properties of the 5-point system
\eqref{5p} from the corresponding (and simpler) properties of the 4-point system
\eqref{4p}.

As it was shown in \cite{NimmoSchief} (see also \cite{DJM}), equation (\ref{4p})
is an integrable discretization (i.e., a discretization possessing
Darboux transformations
(DTs)) of the Moutard equation
\begin{equation}
\label{Moutard}
\Phi_{,uv}=F\Phi
\end{equation}
(the symbol $f_{,u}$ denotes partial differentiation:
$f_{,u}=\partial f/\partial u$).
As it was shown in \cite{NSD},
equation \eqref{5p} is an integrable discretization of the elliptic
(if $AB>1$) equation
\begin{equation}
\label{elliptic}
(A\Phi_{,u})_{,u}+(B\Phi_{,v})_{,v}=F\Phi.
\end{equation}

The paper is organized as follows. In section \ref{sec:connection}
we establish the connection
between the above difference
equations \eqref{4p} and \eqref{5p}.
In sections \ref{sec:DT} and \ref{sec:s-DTs} we use the above connection to
derive the Darboux transformation (DT) and its superposition formula of the
5-point scheme \eqref{5p} (in all its most distinguished gauge equivalent forms)
from the
the Darboux transformation and its superposition formula of the 4-point
scheme \eqref{4p}. In section \ref{sec:al-geom} we
first construct algebro-geometric solutions of the 4-point scheme
\eqref{4p} and then the corresponding solutions of the
5-point scheme \eqref{5p}. Due to the interesting applications of
equations \eqref{4p} and \eqref{5p} (discussed in the second part of this
introduction), these explicit solutions provide examples of discrete holomorphic and harmonic
functions and, at the same time, they generate quadrilateral surfaces in
$\RR^3$.

Let us devote the remaining part of
the introduction to the presentation of the interesting applications
of equations \eqref{4p} and \eqref{5p} in
differential and discrete geometry, and in the
discrete holomorphic function theory.
Equation \eqref{4p} has a very long history. Bianchi constructed the
superposition formula for the Moutard equation (\ref{Moutard})
in the following form \cite{Bianchi}
\begin{equation}
\Phi^{(12)} -\Phi = \frac{\theta^1 \theta^2}{\sigma}
( \Phi^{(1)} - \Phi^{(2)}),
\end{equation}
where $\Phi$, $\Phi^{(1)}$, $\Phi^{(2)}$ and $\Phi^{(12)}$ are four
different general
solutions of four different Moutard equations \eqref{Moutard}, while
$\theta^1$,
$\theta^2$ are particular solutions of the Moutard equation satisfied by
$\Phi$, and $\sigma$ is a potential connected to $\theta^1$, $\theta^2$ via  the
conditions
\begin{align}
\sigma_{,u} & = \theta^1 \theta^2_{,u} - \theta^2 \theta^1_{,u} ,\\
\sigma_{,v} & = \theta^2 \theta^1_{,v} - \theta^1 \theta^2_{,v} .
\end{align}
\begin{figure}
\begin{center}
\mbox{\epsfxsize=6cm \epsffile{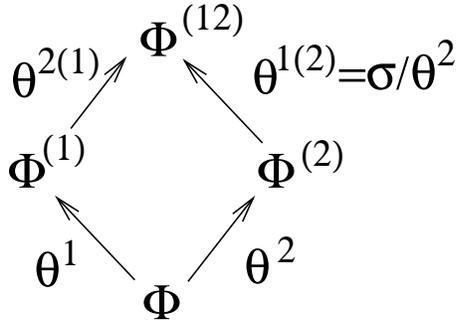}}
\end{center}
\label{fig:M-Bp}
\caption{The Bianchi permutability diagram for the Moutard equation.}
\end{figure}

Using the key observation contained in \cite{LeBen},
such superposition principle was interpreted in
\cite{NimmoSchief} as an integrable discretization \eqref{4p} of the Moutard
equation:
\begin{equation*}
\Phi \to \phi_{m,n}, \quad \Phi^{(1)} \to \phi_{m+1,n}, \quad
\Phi^{(2)} \to \phi_{m,n+1}, \quad \Phi^{(12)} \to \phi_{m+1,n+1}.
\end{equation*}
In addition, DTs and their superposition principle for the
discrete Moutard equation \eqref{4p} were also constructed in \cite{NimmoSchief}.

Solutions of the (real) Moutard equation \eqref{Moutard} allow one to construct,
via the Lelieuvre formulas \cite{Bianchi}, asymptotic nets on hyperbolic
surfaces from their normal vector. Analogously, solutions of the (real)
discrete Moutard equation
\eqref{4p} characterize normal vectors of asymptotic lattices
\cite{Sauer,BP}, which can be constructed via the discrete analogue of the
Lelieuvre formulas \cite{KoPin}.

Also for the elliptic version of equation \eqref{Moutard}, the well known
Schr\"{o}dinger equation
\begin{equation} \label{eq:elliptic-Moutard}
\Psi_{,xx} + \Psi_{,yy} = F\Psi
\end{equation}
(which is equation \eqref{elliptic} with $A=B=1$),
there exists a geometric interpretation within the theory of isothermally
conjugate nets and the corresponding analogue of the Lelieuvre formulas
\cite{Bianchi}. The 5-point scheme \eqref{5p}, introduced in \cite{NSD} as
an integrable discretization of equation \eqref{elliptic}, describes the
normal vector of a quadrilateral lattice in $\RR^3$, and
this embedding is obtained via a suitable generalization of the Lelieuvre
formulas \cite{NS}.

It is remarkable that equations \eqref{4p} and \eqref{5p}, which are basic
equations in the recently developed theory of the Integrable Discrete 
Geometries, are
also the basic equations of an integrable discretization of the theory of
holomorphic and harmonic functions. Indeed the equations
\begin{equation} \label{d-CR}
\psi_{m+1,n+1}-\psi_{m,n}=i(\psi_{m+1,n}-\psi_{m,n+1}),
\end{equation}
\begin{equation} \label{eq:d-harm}
\psi_{\mu+1,\nu}+
\psi_{\mu-1,\nu}+
\psi_{\mu,\nu+1}+
\psi_{\mu,\nu-1}=4
\psi_{\mu,\nu},
\end{equation}
undressed versions of equations \eqref{4p} and \eqref{5p}, were introduced 
as basic objects of
the discrete holomorphic and harmonic function theory in
\cite{Ferrand, Duffin1}; they correspond to a natural discretization of,
respectively, the $\bar\partial$ and Laplace operators on the square lattice,
and the connection between these two schemes was an important ingredient 
of the theory.
In \cite{Duffin2}, this theory was generalized to rhombic lattices; on the 
level of the
4-point scheme, it corresponds to the discrete Cauchy-Riemann equation 
(\ref{4p}) with
a nontrivial pure imaginary potential $g_{m,n}$. In \cite{Mercat} this 
theory was extended to discrete
Riemann surfaces and, in particular, the 4-point scheme (\ref{4p})
and the  affine form
\begin{equation}\label{5p-affine}
a_{\mu,\nu}(\psi_{\mu+1,\nu}-\psi_{\mu,\nu})+
a_{\mu-1,\nu}(\psi_{\mu-1,\nu}-\psi_{\mu,\nu})+
b_{\mu,\nu}(\psi_{\mu,\nu+1}-\psi_{\mu,\nu})+
b_{\mu,\nu-1}(\psi_{\mu,\nu-1}-\psi_{\mu,\nu})=0
\end{equation}
of the self-adjoint 5-point scheme \eqref{5p},
which corresponds to a discretization of the harmonic equation,
were studied on the Riemann surface. A theory of Dirac operators
and the Green's function for planar graphs with rhombic faces were constructed 
in \cite{Kenyon}.
The connection between the above discrete complex function theory and the
discrete complex function theory \cite{Thurston} based on the integrable 
cross-ratio
equation \cite{NijhoffCapel} was established in \cite{BobMerSu}, where
a multidimensional generalization of the discrete complex function theory 
was also considered.

We remark that there exists another natural integrable discretization of 
elliptic
operators on the plane, based on the self-adjoint scheme on the star of
a regular triangular lattice \cite{NovDyn}, resulting in a 7-point scheme. 
One of the
important features of this approach is that this operator is factorized in 
terms of 3-point
operators living on the same lattice. A discrete complex function theory 
based on these operators,
including the discrete analogue of the Cauchy kernel, has been recently 
developed in \cite{DynNov}.

\section{The 5-point scheme as a sub-lattice of the 4-point scheme}
\label{sec:connection}

Consider a lattice, i.e.  a map $\bx :~{\cal D}\to~V$ from a grid $\cal D$
to a linear space $V$
satisfying a certain equation $E[\bx ]=0$ (the lattice equation).
Consider a
subgrid ${\cal D}^\prime\subset{\cal D}$; if one can construct, from the
original lattice
equation $E[\bx ]=0$, a new equation $E'[\bx' ]=0$, where $\bx'$ is the
restriction
of $\bx$ to the sub-grid ${\cal D}^\prime$:
$\bx' =\bx |_{{\cal D}^\prime}:{\cal D}^\prime\to~V$,
then $\bx'$
is a sub-lattice of $\bx$ and $E'[\bx' ]=0$ is the associated
sub-lattice equation.

Suppose that the original lattice (equation) be {\bf integrable}; i.e.,
suppose that one can associate
with it linear transformations enabling one to construct solutions from
solutions (Darboux-type transformations), whose superposition
exhibits permutability properties (the Bianchi permutability diagram).
Since the
infinite class of solutions generated in this way, once restricted,
are also
solutions of the sub-lattice, this sub-lattice (equation) will also be
integrable.
We remark that the dimensional reductions are simple examples of such a
construction.

\subsection{From the lattice to the sub-lattice}
In this section we start with the general 4-point scheme
\begin{equation}
\label{eq:4p-gen}
\alpha_{m,n} \varphi_{m+1,n+1} +\beta_{m,n} \varphi_{m+1,n} + \gamma_{m,n}
\varphi_{m,n+1} +
\delta_{m,n}\varphi_{m,n} =0,
\end{equation}
where $\alpha ,\beta,\gamma$ and $\delta$
are $\CC$-valued functions on ${\cal D}=\ZZ^2$,
and we explore the possibility to construct, from it, a sub-lattice defined on
the
subset $\tilde{\cal D} = \ZZ^2_e$ (the even grid) of $\ZZ^2$ consisting of
points
$(m,n)$ such that $m+n$ is even.
\begin{Prop}
The general 4-point scheme \eqref{eq:4p-gen} reduces to a 5-point scheme
on the even grid if and only if the coefficients $\alpha,\beta,\gamma$ 
and $\delta$ satisfy the constraint
\begin{equation}\label{eq:constr}
\beta_{m,n}\alpha_{m-1,n} \delta_{m,n-1}\gamma_{m-1,n-1} = 
\gamma_{m,n}\delta_{m-1,n}\alpha_{m,n-1}\beta_{m-1,n-1}.
\end{equation}
\end{Prop}
\begin{proof} 
Consider the $4$-point scheme \eqref{eq:4p-gen} in the elementary square
\begin{equation*}
Q_{m,n}=\{(m,n),(m+1,n),(m,n+1),(m+1,n+1)\}
\end{equation*}
and in its
neighbouring squares 
$Q_{m-1,n-1}$, $Q_{m-1,n}$ and $Q_{m,n-1}$,
and move to the LHS 
the terms in which $\varphi$ is evaluated on the odd grid  and to the RHS 
the terms in which $\varphi$ is evaluated on the even grid:
\begin{equation}\label{eq:syst}
\begin{array}{rcl}
\beta_{m,n}\varphi_{m+1,n}+\gamma_{m,n}\varphi_{m,n+1} &=&
-\alpha_{m,n}\varphi_{m+1,n+1}- \delta_{m,n}\varphi_{m,n} ,\\
\alpha_{m-1,n}\varphi_{m,n+1}+\delta_{m-1,n}\varphi_{m-1,n}&=&
-\beta_{m-1,n}\varphi_{m,n} - \gamma_{m-1,n}\varphi_{m-1,n+1}, \\
\alpha_{m,n-1}\varphi_{m+1,n} +\delta_{m,n-1}\varphi_{m,n-1}&=&
-\gamma_{m,n-1}\varphi_{m,n} - \beta_{m,n-1}\varphi_{m+1,n-1}, \\
\gamma_{m-1,n-1}\varphi_{m-1,n} + \beta_{m-1,n-1}\varphi_{m,n-1}&=&
-\alpha_{m-1,n-1}\varphi_{m,n} - \delta_{m-1,n-1}\varphi_{m-1,n-1}. 
\end{array}
\end{equation}
In order to construct a lattice equation not involving $\varphi$ on the odd 
grid, it is
sufficient to impose the zero determinant condition for the system
\eqref{eq:syst}, which coincides with the constraint \eqref{eq:constr}. 
\end{proof} 
\begin{Rem}
The row vector solution $\bv$ of the homogeneous version of the system 
\eqref{eq:syst} reads:
\begin{equation}
\begin{split}
{\bv}_{m,n} = & (
\gamma_{m-1,n-1}\alpha_{m-1,n}\alpha_{m,n-1},
-\gamma_{m,n}\gamma_{m-1,n-1} \alpha_{m,n-1} , \\
&-\alpha_{m-1,n}\beta_{m,n} \gamma_{m-1,n-1} , 
\alpha_{m,n-1}\delta_{m-1,n}\gamma_{m,n} )
\end{split}
\end{equation}
and the wanted sub-lattice equation is the solvability
condition $({\bv}_{m,n},{\bF}_{m,n})=0$ of the Kronecker - Capelli theorem, 
where
\begin{equation}
\begin{array}{l}
\bF_{m,n}=-(\alpha_{m,n}\varphi_{m+1,n+1}+\delta_{m,n}\varphi_{m,n},\; 
\gamma_{m-1,n}\varphi_{m-1,n+1} +\beta_{m-1,n}\varphi_{m,n}, \\
\beta_{m,n-1}\varphi_{m+1,n-1} +\gamma_{m,n-1}\varphi_{m,n}, \;
\delta_{m-1,n-1}\varphi_{m-1,n-1} + \alpha_{m-1,n-1}\varphi_{m,n})^T
\end{array}
\end{equation}
is the column vector representing the inhomogeneous term of the algebraic 
system.
This solvability condition reads:
\begin{equation}
\label{eq:5pbis}
\begin{array}{l}
\gamma_{m-1,n-1}\alpha_{m-1,n}\alpha_{m,n-1} 
(\alpha_{m,n}\varphi_{m+1,n+1}+\delta_{m,n}\varphi_{m,n}) - \\
\gamma_{m,n}\gamma_{m-1,n-1} \alpha_{m,n-1}
(\gamma_{m-1,n}\varphi_{m-1,n+1} +\beta_{m-1,n}\varphi_{m,n}) -\\
\alpha_{m-1,n}\beta_{m,n} \gamma_{m-1,n-1}
(\beta_{m,n-1}\varphi_{m+1,n-1} +\gamma_{m,n-1}\varphi_{m,n}) +\\
\alpha_{m,n-1}\delta_{m-1,n}\gamma_{m,n} 
(\delta_{m-1,n-1}\varphi_{m-1,n-1} + \alpha_{m-1,n-1}\varphi_{m,n})
=0.
\end{array}
\end{equation}
\end{Rem}

\begin{Prop} \label{prop:4p-can}
The four point scheme \eqref{eq:4p-gen} with non-vanishing coefficients 
$\alpha,\beta,\gamma, \delta$ satisfying constraint
\eqref{eq:constr} is gauge equivalent to the discrete Moutard (discrete
Cauchy--Riemann) four point scheme \eqref{4p} whose corresponding five
point scheme is the affine self-adjoint five point scheme \eqref{5p-affine}.
\end{Prop}
\begin{proof}
Let the potential $\rho :\ZZ^2\to\CC$ be a solution of the system
\begin{equation}\label{eq:param}
\alpha_{m,n} \rho_{m+1,n+1} = - \delta_{m,n}\rho_{m,n},
\qquad \beta_{m,n}\rho_{m+1,n}  = - \gamma_{m,n}\rho_{m,n+1},
\end{equation}
whose compatibility condition is the constraint \eqref{eq:constr}.
Then the function $\psi:\ZZ^2\to\CC$
\begin{equation}\label{eq:gauge}
\psi_{m,n} =\varphi_{m,n} /\rho_{m,n} ,
\end{equation}
satisfies equation 
\begin{equation}
\label{4p-affine}
\psi_{m+1,n+1}-\psi_{m,n}=if_{m,n}(\psi_{m+1,n}-\psi_{m,n+1}),
\end{equation}
with the potential
\begin{equation}
g_{m,n}=if_{m,n}=\frac{\beta_{m,n}\rho_{m+1,n}}{\delta_{m,n}\rho_{m,n}}=
-\frac{\gamma_{m,n}\rho_{m,n+1}}{\delta_{m,n}\rho_{m,n}},
\end{equation}
and equation \eqref{eq:5pbis} reduces to the 5-point scheme
\begin{equation}
\label{5p-rot}
\begin{array}{l}
a_{m,n}(\psi_{m+1,n-1}-\psi_{m,n})+
a_{m-1,n+1}(\psi_{m-1,n+1}-\psi_{m,n})+\\
b_{m,n}(\psi_{m+1,n+1}-\psi_{m,n})+
b_{m-1,n-1}(\psi_{m-1,n-1}-\psi_{m,n})=0,
\end{array}
\end{equation}
with the functions $a_{m,n}$ and $b_{m,n}$ defined as follows
\begin{equation}
\label{eq:f-ab}
a_{m,n} = f_{m,n-1}, \qquad b_{m,n}=\frac{1}{f_{m,n}} .
\end{equation}
The connection between \eqref{4p-affine} and \eqref{5p-rot}
we have established here can be easily verified directly considering
the $4$-point scheme
(\ref{4p-affine}) for the vector $\psi \in V$ in the elementary square
$Q_{m,n}$ and in its
neighbouring squares $Q_{m-1,n-1}$, $Q_{m-1,n}$, and $Q_{m,n-1}$:
\begin{equation}
\label{4p-4}
\begin{array}{rcr}
\frac{1}{f_{m,n}}(\psi_{m+1,n+1}-\psi_{m,n})& = 
&i(\psi_{m+1,n}-\psi_{m,n+1}), \\
\frac{1}{f_{m-1,n-1}}(\psi_{m-1,n-1}-\psi_{m,n})& =
& i(\psi_{m-1,n}-\psi_{m,n-1}), \\
f_{m,n-1}(\psi_{m+1,n-1}-\psi_{m,n})& = & -i(\psi_{m+1,n}-\psi_{m,n-1}), \\
f_{m-1,n}(\psi_{m-1,n+1}-\psi_{m,n})& =&i(\psi_{m,n+1}-\psi_{m-1,n}).
\end{array}
\end{equation}
Adding these four equations up, one obtains the 5-point scheme \eqref{5p-rot}.

\begin{figure}
\begin{center}
\leavevmode\epsfxsize=10cm \epsffile{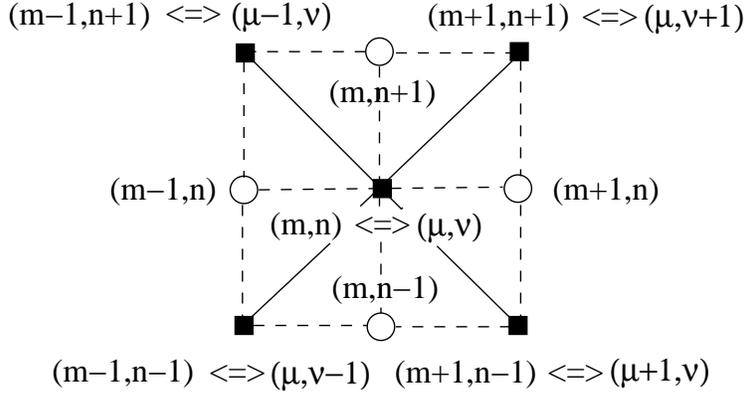}
\label{fig:4P-5P}
\caption{The white (odd) points are eliminated taking a suitable linear
combination
of four adjacent 4-point schemes. What remains is a $\pi /4$ rotated
5-point scheme
on the black (even) points.}
\end{center}
\end{figure}
We are therefore lead to the following change of variables in $\ZZ^2_e$ (see
figure 2):
\begin{equation} \label{eq:mn-mn}
\mu = \frac{m-n}{2}, \qquad \nu = \frac{n+m}{2},
\end{equation}
corresponding to a $\pi /4$ rotation of the axes and, in the new variables,
equation \eqref{5p-rot} is the affine self-adjoint five point scheme
\eqref{5p-affine}.
\end{proof}
\begin{Rem}
Notice that the potential $\rho$ satisfies the constrained
equation \eqref{eq:4p-gen}
\end{Rem}
\begin{Rem}
Any self-adjoint 5-point scheme \eqref{5p}
is gauge-equivalent to an affine self-adjoint 5-point scheme \eqref{5p-affine}.
Indeed, if $\sigma_{\mu,\nu}$ is a solution of \eqref{5p} with coefficients
$a_{\mu,\nu}$, $b_{\mu,\nu}$ and $c_{\mu,\nu}$, then 
\begin{equation*}
\psi_{\mu,\nu}= \Psi_{\mu,\nu}/\sigma_{\mu,\nu}
\end{equation*}
satisfies the affine equation \eqref{5p-affine} with coefficients
\begin{equation*}
a_{\mu,\nu}^{\text{aff}} = a_{\mu,\nu}\, \sigma_{\mu+1,\nu} \, \sigma_{\mu,\nu},\qquad
b_{\mu,\nu}^{\text{aff}} = b_{\mu,\nu} \, \sigma_{\mu,\nu+1} \, 
\sigma_{\mu,\nu}.
\end{equation*} 
\end{Rem}

\subsection{From the sub-lattice to the lattice}
\label{sec:subl-l}
The results of the previous section 2.1 leave open the question if
all the solutions $\psi : \ZZ^2_e \to \CC$ of the 5-point
scheme (\ref{5p-rot}) can be extended to solutions of the 4-point scheme
(\ref{4p-affine})
on the whole lattice. The answer is affirmative and the construction is
very simple.

Suppose one knows a solution $\psi :\ZZ^2_e \to \CC$ of the
5-point scheme (\ref{5p-rot}) for some given coefficients
$a,b : \ZZ^2_e \to \RR$.
Define the function $f:\ZZ^2 \to \RR$ on the whole lattice by the 
inverses of
equations \eqref{eq:f-ab} and \eqref{eq:mn-mn},
and choose the value of $\psi$ at one arbitrary odd point of the lattice.
Then, using the 4-point scheme (\ref{4p-affine}) and the known values of 
$\psi$ at the even points, one can construct uniquely a solution  $\psi$
of the 4-point scheme (\ref{4p-affine}) whose restriction to $\ZZ^2_e$
coincides with
$\psi$.

Summarizing the content of the last two sections, we have shown that 
any solution of the
4-point scheme (\ref{4p}), once restricted to $\ZZ^2_e$, generates a
solution of the 5-point scheme (\ref{5p-rot}). Viceversa, {\bf any} solution
of the
5-point scheme (\ref{5p-rot}) can be
obtained restricting to $\ZZ^2_e$ a suitable solution of the 4-point scheme
(\ref{4p-affine}).

\subsection{Discretization of the Schr\"{o}dinger equation via the Moutard
transformation}
In this section we show that also the method of
"discretization via transformations"
\cite{LeBen}, when applied to the elliptic Moutard (Schr\"{o}dinger) 
equation
\eqref{eq:elliptic-Moutard}, leads to the 5-point scheme thanks
to the sublattice approach.
Consider the general solution $\Psi$ of the Schr\"{o}dinger equation
\eqref{eq:elliptic-Moutard} together with its particular solutions 
$\theta^1$ and $\theta^2$. One can easily check that the functions
$\Psi^{(j)}$, $j=1,2$, given as solutions of the compatible linear systems
\begin{align}
(\theta^j \Psi^{(j)})_{,x} & = -\theta^j \Psi_{,y} + \theta^j_{,y} \Psi, \\
(\theta^j \Psi^{(j)})_{,y} & = \; \; \theta^j \Psi_{,x} - \theta^j_{,y} \Psi,
\end{align}
satisfy again the Schr\"{o}dinger equations
\eqref{eq:elliptic-Moutard}, but with the new potentials
\begin{equation}
F^{(j)} = \theta^j \left[ \left(\frac{1}{\theta^j}  \right)_{,xx} +
\left(\frac{1}{\theta^j}  \right)_{,yy} \right].
\end{equation}
Denote by $\sigma$ a solution of the system
\begin{align}
\sigma_{,x} & = - \theta^1 \theta^2_{,y} + \theta^1_{,y} \theta^2  ,\\
\sigma_{,y} & = \; \; \theta^1\theta^2_{,x} - \theta^1_{,x} \theta^2 ,
\end{align}
then the functions $\theta^{1(2)}$ and $\theta^{2(1)}$ given by
\begin{equation}
\sigma = \theta^2\theta^{1(2)} = - \theta^1\theta^{2(1)},
\end{equation}
satisfy the same equations as, respectively, $\Psi^{(2)}$ and $\Psi^{(1)}$. 
Finally, the function $\Psi^{(12)}$, obtained from the
superposition formula
\begin{equation} \label{eq:sup-M-el}
\Psi^{(12)} +\Psi = \frac{\theta^1 \theta^2}{\sigma}
( \Psi^{(2)} - \Psi^{(1)}),
\end{equation}
satisfies the Schr\"{o}dinger equation \eqref{eq:elliptic-Moutard} with
potential
\begin{equation}
F^{(12)} = F - \frac{2}{\sigma}\left( \theta^1_{,x} \theta^2_{,y} -
\theta^1_{,y} \theta^2_{,x}    \right) +
\sigma \left[ \left(\frac{1}{\sigma }  \right)_{,xx} +
\left(\frac{1}{\sigma }  \right)_{,yy} \right],
\end{equation}
and is simultaneously the transform of $\Psi^{(1)}$ via $\theta^{2(1)}$,
and the transform of $\Psi^{(2)}$ via $\theta^{1(2)}$.

Therefore the superposition principle of the Schr\"{o}dinger
equation \eqref{eq:elliptic-Moutard} is described by the 4-point scheme 
\eqref{eq:sup-M-el} of the discrete Moutard type, which we know is not 
a proper
discretization of \eqref{eq:elliptic-Moutard}. This seems to be in contradiction
with the general rule that the superposition principle of a continuous system
provides an integrable discretization of it. However, as we know, 
the sub-lattice approach resolves 
this problem, since the 4-point scheme \eqref{eq:sup-M-el} reduces, on its
subgrid, to the 5-point scheme (see also section \ref{sec:gauges}), 
which turns out to be the proper discrete
analogue of \eqref{eq:elliptic-Moutard} \cite{NSD}.

\subsection{Different gauge forms of the 5-point scheme}
\label{sec:gauges}
For the construction of the Darboux transformations and in other applications
it is useful
to introduce the $\tau$-function of the 4-point scheme \eqref{4p-affine}
\begin{equation} \label{eq:tau}
\frac{\tau_{m+1,n}\tau_{m,n+1}}{\tau_{m,n}\tau_{m+1,n+1}} =f_{m,n}.
\end{equation}
The fact presented below, which can be checked by direct calculation, 
explains the introduction of
such a potential in a broader context.
\begin{Th}
Consider a lattice $\bpsi:\ZZ^N\to V$, $\dim V\geq N\geq 3$, satisfying the
following set of linear problems
\begin{equation} \label{eq:lin_BKP}
\bpsi_{m..(i+1)..(j+1)..n} - \bpsi_{m..i..j..n} = g_{m..i..j..n}^{ij}
(\bpsi_{m..(i+1)..j..n} - \bpsi_{m..i..(j+1)..n}), \quad i< j ,
\end{equation}
then the functions $g_{m..i..j..n}^{ij}$ can be parametrized by the potential
$\tau_{m..i..j..n}$
\begin{equation*}
g_{m..i..j..n}^{ij} = \frac{\tau_{m..(i+1)..j..n}\tau_{m..i..(j+1)..n}}
{\tau_{m..(i+1)..(j+1)..n}\tau_{m..i..j..n}}, \qquad i<j,
\end{equation*}
and the compatibility condition of the linear system \eqref{eq:lin_BKP}
gives the nonlinear system of discrete BKP equations \cite{Miwa}
\begin{multline*}
\tau_{m..i..j..k..n} \tau_{m..(i+1)..(j+1)..(k+1)..n} -
\tau_{m..(i+1)..j..k..n} \tau_{m..i..(j+1)..(k+1)..n} + \\
\tau_{m..i..(j+1)..k..n} \tau_{m..(i+1)..j..(k+1)..n} - 
\tau_{m..i..j..(k+1)..n} \tau_{m..(i+1)..(j+1)..k..n} =0, 
\quad i<j<k.
\end{multline*}
\end{Th} 

Let us define the function
\begin{equation} \label{eq:def-Phi}
\Phi_{m,n} = \frac{\tau_{m,n}}{\tau_{m+1,n}} \psi_{m,n},
\end{equation}
then, by direct calculation using equations \eqref{5p-rot} and 
\eqref{eq:f-ab}, one
can check that the function $\Phi$, restricted to the even grid $\ZZ_e^2$,
satisfies the equation
\begin{equation} \label{eq:5p-eq-f}
h_{\mu+1,\nu}\Phi_{\mu+1,\nu}+
h_{\mu,\nu}\Phi_{\mu-1,\nu}+
h_{\mu,\nu+1}\Phi_{\mu,\nu+1}+
h_{\mu,\nu}\Phi_{\mu,\nu-1}=c_{\mu,\nu} \Phi_{\mu,\nu},
\end{equation}
where
\begin{equation} \label{eq:def-h}
h_{m,n} = \frac{\tau_{m+1,n}}{\tau_{m-1,n}},
\end{equation}
and
\begin{equation} \label{eq:def-c}
c_{m,n} =\frac{\tau_{m+1,n}}{\tau_{m,n}}
\left( \frac{\tau_{m+1,n-1}}{\tau_{m,n-1}} +
\frac{\tau_{m+1,n+1}}{\tau_{m,n+1}} \right) + 
\frac{\tau_{m+1,n}^2}{\tau_{m,n}\tau_{m-1,n}} 
\left( \frac{\tau_{m-1,n+1}}{\tau_{m,n+1}} +
\frac{\tau_{m-1,n-1}}{\tau_{m,n-1}} \right) .
\end{equation}
We call equation \eqref{eq:5p-eq-f} the equal-field self-adjoint 5-point 
scheme.

If we define the function
\begin{equation} \label{eq:gauge-5p-aff-Sch}
\Psi_{m,n} = \frac{\tau_{m,n}}{\sqrt{\tau_{m+1,n}\tau_{m-1,n}}} \psi_{m,n},
\end{equation}
then the function $\Psi$, restricted to the even grid $\ZZ_e^2$,
satisfies the discrete Schr\"{o}dinger equation \cite{NSD}
\begin{equation} \label{eq:Sch}
\frac{\Gamma_{\mu,\nu}}{\Gamma_{\mu+1,\nu}}\Psi_{\mu+1,\nu}+
\frac{\Gamma_{\mu-1,\nu}}{\Gamma_{\mu,\nu}}\Psi_{\mu-1,\nu}+
\frac{\Gamma_{\mu,\nu}}{\Gamma_{\mu,\nu+1}}\Psi_{\mu,\nu+1}+
\frac{\Gamma_{\mu,\nu-1}}{\Gamma_{\mu,\nu}}\Psi_{\mu,\nu-1}=
F_{\mu,\nu} \Psi_{\mu,\nu},
\end{equation}
where
\begin{equation} \label{eq:Sch-Gamma}
\Gamma_{m,n} = \sqrt{\frac{\tau_{m-1,n}}{\tau_{m+1,n}}},
\end{equation}
and
\begin{equation} \label{eq:def-F}
F_{m,n} =
\frac{\tau_{m-1,n}}{\tau_{m,n}}
\left( \frac{\tau_{m+1,n-1}}{\tau_{m,n-1}} +
\frac{\tau_{m+1,n+1}}{\tau_{m,n+1}} \right) + 
\frac{\tau_{m+1,n}}{\tau_{m,n}} 
\left( \frac{\tau_{m-1,n+1}}{\tau_{m,n+1}} +
\frac{\tau_{m-1,n-1}}{\tau_{m,n-1}} \right)  .
\end{equation}
Notice the following connection formulas between the equal-field and the
Schr\"{o}dinger gauges:
\begin{equation} \label{eq:con-ef-S}
\Phi_{\mu,\nu}  = \Gamma_{\mu,\nu}\Psi_{\mu,\nu}, \qquad
h_{\mu,\nu}  = \frac{1}{\Gamma_{\mu,\nu}^2},\qquad
c_{\mu,\nu} = \frac{1}{\Gamma_{\mu,\nu}^2} F_{\mu,\nu}.
\end{equation}
\begin{Rem}
The potential $\tau$ is defined up to multiplication by functions of single 
variables $m$ and $n$, which implies that also $\Phi$ and $\Psi$ are not unique.
\end{Rem}
Finally we mention a direct consequence of the above definitions of the
equal-field and the Schr\"{o}dinger gauges.
\begin{Cor}
Given a solution $\theta:\ZZ^2\to\CC$ of the discrete Moutard equation
\eqref{4p-affine} with the $\tau$-function $\tau_{m,n}$, then\\
a) $\theta$ restricted to the sub-lattice satisfies the affine 5-point scheme 
\eqref{5p-affine};\\
b) the function
\begin{equation} \label{eq:def-rho}
\rho_{m,n} =\frac{\tau_{m,n}}{\tau_{m+1,n}} \theta_{m,n},
\end{equation}
restricted to the 
sub-lattice, satisfies the corresponding equal-field 5-point scheme
\eqref{eq:5p-eq-f};\\
c) the function
\begin{equation} \label{eq:def-Theta}
\Theta_{m,n} = \frac{\tau_{m,n}}{\sqrt{\tau_{m+1,n}\tau_{m-1,n}}} 
\theta_{m,n} = \frac{1}{\Gamma_{m,n}}\rho_{m,n},
\end{equation}
restricted to the 
sub-lattice, satisfies the corresponding Schr\"{o}dinger 5-point scheme
\eqref{eq:Sch}.
\end{Cor}

\section{Darboux transformations of the sub-lattice}
\label{sec:DT}
An obvious application of the above construction is that,
once a sub-lattice of a given integrable lattice is identified , one
obtains
essentially for free some of its integrability properties.
Here we show, for instance,
the construction of the Darboux transformations \cite{NSD}
of the 5-point scheme in the affine \eqref{5p-affine}, 
equal-field \eqref{eq:5p-eq-f}
and Schr\"odinger \eqref{eq:Sch} forms, 
induced by the transformations of the 4-point lattice \eqref{4p-affine}.

\subsection{DTs of the affine 5-point scheme}
\label{sec:DT-aff5p}
We first recall relevant material~\cite{NimmoSchief} 
on the Darboux-Moutard transformations
of the  4-point lattice.
\begin{Prop} Given a
solution $\theta$ of the discrete 4-point scheme \eqref{4p-affine}, then
any solution $\tilde\psi$
of the compatible linear system of the first order
\begin{align} \label{eq:Mt1}
\tilde\psi_{m+1,n} + \psi_{m,n} & =
\frac{\theta_{m,n}}{\theta_{m+1,n}}(\psi_{m+1,n} + \tilde\psi_{m,n}), \\
\label{eq:Mt2}
\tilde\psi_{m,n+1}  + \psi_{m,n} & =
\frac{\theta_{m,n}}{\theta_{m,n+1}}(\psi_{m,n+1} + \tilde\psi_{m,n}),
\end{align}
satisfies the 4-point scheme \eqref{4p-affine} with the transformed
potential
\begin{equation} \label{eq:tilde-F}
\tilde{f}_{m,n} = \frac{\theta_{m+1,n} \theta_{m,n+1}}
{\theta_{m,n} \theta_{m+1,n+1}}f_{m,n},
\end{equation}
while the corresponding transformation of the $\tau$-function takes
the simple form
\begin{equation} \label{eq:tilde-tau}
\tilde\tau_{m,n} = \theta_{m,n}\tau_{m,n}.
\end{equation}
\end{Prop}
\begin{Rem}
Notice that the function 
\begin{equation} \tilde\theta_{m,n} = \frac{1}{\theta_{m,n}}
\end{equation}
is a solution the 4-point scheme \eqref{4p-affine} of $\tilde\psi$.
\end{Rem}
\begin{Cor} \label{cor:DT-int-3d}
One can interpret the transformation as a shift in the third dimension of the
lattice, and the transformation equations \eqref{eq:Mt1}-\eqref{eq:Mt2} as
linear problems of the form \eqref{4p-affine} involving that dimension. In
particular, the transformation rule of the $\tau$-function
\eqref{eq:tilde-tau} and the parametrization \eqref{eq:tau} of the potential
$f_{m,n}$ in equation \eqref{4p-affine} satisfied by $\theta_{m,n}$ lead to
equation 
\begin{equation}
\tau_{m,n} \tilde\tau_{m+1,n+1} - \tilde\tau_{m,n}\tau_{m+1,n+1} =
i(\tau_{m,n+1} \tilde\tau_{m+1,n} - \tau_{m+1,n}\tilde\tau_{m,n+1}),
\end{equation}
of the form of the discrete BKP equation for $N=3$.
\end{Cor}

Again, the function $\tilde\psi_{m,n}$, when restricted to the even (or odd)
lattice, satisfies the affine 5-point scheme \eqref{5p-affine}. However, the
transformation equations \eqref{eq:Mt1}-\eqref{eq:Mt2} depend on values of
the transformation potential $\theta$ on the full lattice. Our goal is to
constrain the DT of the 5-point scheme to the sub-lattice as well. Obviously,
given a solution of the affine 5-point scheme on the sub-lattice, it can be 
propagated to the full lattice in the spirit of section~\ref{sec:subl-l} and
then used to construct the transformation. We will show, however, that the
transformations of the 5-point scheme can be done, in a more elegant way.
In particular we will constrain the transformation equations to the sub-lattice.
\begin{Lem}
Let $\psi_{m,n}$ and $\theta_{m,n}$ be solutions of \eqref{4p-affine}, and let
$\tilde\psi_{m,n}$ be the transformed solution constructed via 
\eqref{eq:Mt1}-\eqref{eq:Mt2}. Then the function  $\hat\psi_{m,n} $, defined by
\begin{equation} \label{eq:def-h-psi}
\hat\psi_{m,n} = i\theta_{m-1,n}\tilde\psi_{m-1,n},
\end{equation}
with the functions $a_{m,n}$ and $b_{m,n}$ defined by equations \eqref{eq:f-ab},
satisfies
\begin{align} \label{eq:4-5-DT-L1}
\hat\psi_{m+1,n-1} - \hat\psi_{m,n} & =
b_{m-1,n-1}(\theta_{m-1,n+1}\psi_{m,n} -\theta_{m,n}\psi_{m-1,n+1}),\\
\label{eq:4-5-DT-L2}
\hat\psi_{m+1,n+1} -\hat\psi_{m,n}& =
-a_{m-1,n+1}(\theta_{m-1,n+1}\psi_{m,n} -
\theta_{m,n}\psi_{m-1,n+1}).
\end{align}
\end{Lem}
\begin{proof}
Substract equation \eqref{eq:Mt2} from equation \eqref{eq:Mt1}, use the four
point scheme \eqref{4p-affine} and evaluate the result at $(m-1,n-1)$ to get
equation \eqref{eq:4-5-DT-L1}. Similarly, add equation \eqref{eq:Mt1} 
evaluated at
$(m-1,n)$ to
equation \eqref{eq:Mt2} to obtain equation \eqref{eq:4-5-DT-L2}.
\end{proof}

\begin{Cor}
The function $\hat\psi_{m,n}$
satisfies equation
\begin{equation}
\begin{array}{l}
\hat{a}_{m,n}(\hat\psi_{m+1,n-1}-\hat\psi_{m,n})+
\hat{a}_{m-1,n+1}(\hat\psi_{m-1,n+1}-\hat\psi_{m,n})+\\
\hat{b}_{m,n}(\hat\psi_{m+1,n+1}-\hat\psi_{m,n})+
\hat{b}_{m-1,n-1}(\hat\psi_{m-1,n-1}-\hat\psi_{m,n})=0,
\end{array}
\end{equation}
with coefficients
\begin{equation} \label{eq:4-5-DT-hab}
\hat{a}_{m,n}  = \frac{1}{b_{m-1,n-1}\theta_{m,n}\theta_{m-1,n-1}},\qquad
\hat{b}_{m,n}  = \frac{1}{a_{m-1,n+1}\theta_{m,n}\theta_{m-1,n+1}}.
\end{equation}
\end{Cor}
\begin{proof}
The function  $\tilde\psi_{m,n}$ satisfies equation \eqref{5p-rot}
with coefficients $\tilde{a}_{m,n}$ and $\tilde{b}_{m,n}$ obtained
from $\tilde{f}_{m,n}$
via equations \eqref{eq:f-ab}.
Because the function
$1/\theta_{m,n}$ satisfies the same equation as $\tilde\psi_{m,n}$, then
$\theta_{m,n}\tilde\psi_{m,n}$
satisfies again equation \eqref{5p-rot}, but with new coefficients (compare with
remark 2 after proposition \ref{prop:4p-can}).
\end{proof}
The following theorem can be obtained just restricting the above results to the
sub-lattice, but it can be also proven directly.
\begin{Th} \label{th:DT-5-aff}
Let $\psi_{\mu,\nu}$ satisfy the self-adjoint affine 5-point scheme
\eqref{5p-affine} and let $\theta_{\mu,\nu}$ be a particular solution of
\eqref{5p-affine}. Then the solution $\hat\psi_{\mu,\nu}$ of the compatible
linear system
\begin{align}
\label{eq:5-DT-aff1}
\hat\psi_{\mu+1,\nu} - \hat\psi_{\mu,\nu} & =
\; b_{\mu,\nu-1}(\theta_{\mu,\nu-1}\psi_{\mu,\nu} -
\theta_{\mu,\nu}\psi_{\mu,\nu-1}),\\
\label{eq:5-DT-aff2}
\hat\psi_{\mu,\nu+1} - \hat\psi_{\mu,\nu} & =
- a_{\mu-1,\nu}(\theta_{\mu-1,\nu}\psi_{\mu,\nu} -
\theta_{\mu,\nu}\psi_{\mu-1,\nu}),
\end{align}
satisfies the self-adjoint affine 5-point scheme
\eqref{5p-affine} with transformed coefficients
\begin{equation} \label{eq:5-DT-aff-hab}
\hat{a}_{\mu,\nu} =
\frac{1}{b_{\mu,\nu-1}\theta_{\mu,\nu}\theta_{\mu,\nu-1}},\qquad
\hat{b}_{\mu,\nu} =
\frac{1}{a_{\mu-1,\nu}\theta_{\mu,\nu}\theta_{\mu-1,\nu}}.
\end{equation}
\end{Th}
\begin{proof}
The compatibility condition of
\eqref{eq:5-DT-aff1}-\eqref{eq:5-DT-aff2} is the self-adjoint affine 5-point
scheme \eqref{5p-affine}.
Equations \eqref{eq:5-DT-aff1}-\eqref{eq:5-DT-aff2} and their compatibility
condition imply the self-adjoint affine 5-point scheme
\eqref{5p-affine} for $\hat\psi_{\mu,\nu}$
with coefficients given by \eqref{eq:5-DT-aff-hab}.
\end{proof}
\begin{Rem}
The transformation \eqref{eq:5-DT-aff1}-\eqref{eq:5-DT-aff2} is a restriction
of the DT of the general self-adjoint 5-point scheme given in
\cite{NSD} to the class
of self-adjoint affine 5-point schemes.
\end{Rem}
\begin{Rem}
Equations \eqref{eq:5-DT-aff1}-\eqref{eq:5-DT-aff2} are obtained
restricting equations \eqref{eq:4-5-DT-L1}-\eqref{eq:4-5-DT-L2} to the
sub-lattice, while equation \eqref{eq:5-DT-aff-hab} is the corresponding
restriction of equations \eqref{eq:4-5-DT-hab}.
\end{Rem}

\begin{figure}
\begin{center}
\mbox{\epsfxsize=8cm \epsffile{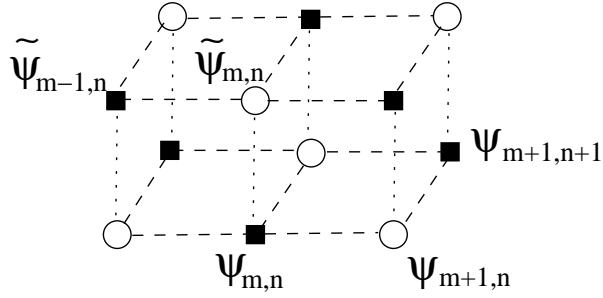}}
\end{center}
\label{fig:DT-red}
\caption{The sub-lattice reduction of the discrete Moutard transformation.}
\end{figure}

It seems that the 
transformation
acts from the even (black) lattice $\ZZ_e^2$ to the odd (white)
lattice $\ZZ_o^2$ (or vice
versa). However, the proper point of view is to consider 
(see corollary \ref{cor:DT-int-3d}) the transformation as
the shift into an additional dimension of the lattice, and then 
the transformation acts between black (or between white) points 
of the lattice
(see figure 3). 
There are three other equivalent choices of the transformation on the small
lattice which give also restrictions of the DT to the
sub-lattice. The resulting transformation formulas differ from
\eqref{eq:5-DT-aff1}-\eqref{eq:5-DT-aff2} in some shifts. The present form
has been chosen to be in agreement with the earlier formulas of \cite{NSD}.

\subsection{DTs of the self-adjoint equal field 5-point scheme}
Let us rewrite equations \eqref{eq:4-5-DT-L1}-\eqref{eq:4-5-DT-L2} 
using, instead of
the fields $\psi_{m,n}$, $a_{m,n}$, $b_{m,n}$ and $\theta_{m,n}$, the fields
$\Phi_{m,n}$, $h_{m,n}$ and $\rho_{m,n}$ defined by equations \eqref{eq:def-Phi}, 
\eqref{eq:def-h} and \eqref{eq:def-rho}. The natural counterpart of the
function $\hat\psi_{m,n}$ given by equation \eqref{eq:def-h-psi} is the function
\begin{equation} \label{eq:def-h-Phi}
\hat{\Phi}_{m,n} = i \tilde\Phi_{m-1,n},
\end{equation}
where $\tilde\Phi_{m,n}$ is given by the transformed version of equation
\eqref{eq:def-Phi}, i.e., 
\begin{equation} \label{eq:def-t-Phi}
\tilde\Phi_{m,n} =
\frac{\tilde\tau_{m,n}}{\tilde\tau_{m+1,n}}\tilde\psi_{m,n} =
\frac{\theta_{m,n}\tilde\psi_{m,n}}{\rho_{m+1,n}h_{m+1,n}}.
\end{equation}
The counterpart of the DT of the affine self-adjoint 5-point scheme described in
theorem \ref{th:DT-5-aff} will be the following theorem on the DT of the
self-adjoint equal field 5-point scheme.
\begin{Th} \label{th:DT-ef}
Let $\Phi_{\mu,\nu}$ and $\rho_{\mu,\nu}$ be solutions of the self-adjoint
equal field 5-point scheme \eqref{eq:5p-eq-f}, then the solution
$\hat\Phi_{\mu,\nu}$ of the following linear system of the first order
\begin{align}
\rho_{\mu+1,\nu}h_{\mu+1,\nu}\hat\Phi_{\mu+1,\nu} -
\rho_{\mu,\nu}h_{\mu,\nu}\hat\Phi_{\mu,\nu}  & =
\; \; h_{\mu,\nu} \left(
\rho_{\mu,\nu-1}\Phi_{\mu,\nu} - \rho_{\mu,\nu}\Phi_{\mu,\nu-1}
\right) ,\\
\rho_{\mu,\nu+1}h_{\mu,\nu+1}\hat\Phi_{\mu,\nu+1} -
\rho_{\mu,\nu}h_{\mu,\nu}\hat\Phi_{\mu,\nu}  & =
- h_{\mu,\nu} \left(
\rho_{\mu-1,\nu}\Phi_{\mu,\nu} - \rho_{\mu,\nu}\Phi_{\mu-1,\nu}
\right) ,
\end{align}
satisfies the self-adjoint equal field 5-point scheme with coefficients
\begin{equation} \label{eq:def-h-h}
\hat{h}_{\mu,\nu} = \frac{h_{\mu,\nu}\rho_{\mu,\nu}}{\rho_{\mu-1,\nu-1}},
\end{equation}
and
\begin{equation} \label{eq:c-hat}
\hat{c}_{\mu,\nu} = h_{\mu,\nu}\rho_{\mu,\nu}
\left( \frac{1}{\rho_{\mu,\nu-1}} +
\frac{1}{\rho_{\mu-1,\nu}} \right) +
\frac{(h_{\mu,\nu}\rho_{\mu,\nu})^2}
{\rho_{\mu-1,\nu-1}} \left(
\frac{1}{h_{\mu-1,\nu}\rho_{\mu-1,\nu}}+  
\frac{1}{h_{\mu,\nu-1}\rho_{\mu,\nu-1}} 
\right).
\end{equation}
\end{Th}
\begin{proof}
The theorem can be verified by direct calculation. To do its derivation
within the sub-lattice theory (the form of the 
transformation equations has been explained  
before the formulation of the theorem), notice that the function
$\hat{h}_{\mu,\nu}$ is restriction of 
\begin{equation} \label{eq:def-t-h}
\hat{h}_{m,n}=\tilde{h}_{m-1,n}, \qquad \text{with} \quad
\tilde{h}_{m,n} = \frac{\tilde\tau_{m+1,n}}{\tilde\tau_{m-1,n}}.
\end{equation} 
Also the transformed coefficient $\hat{c}_{\mu,\nu}$ is the restriction of 
\begin{equation}
\hat{c}_{m,n} = \tilde{c}_{m-1,n}, 
\end{equation}
with $\tilde{c}_{m,n}$ given by equation \eqref{eq:def-c}, with
$\tilde\tau_{m,n}$ instead of $\tau_{m,n}$. In putting it on the
sub-lattice one uses
equation \eqref{4p-affine} satisfied by $\theta_{m,n}$. 
\end{proof}
\begin{Rem}
In terms of the function
\begin{equation} \label{eq:def-h-rho}
\hat{\rho}_{\mu,\nu} = \frac{1}{\rho_{\mu,\nu}h_{\mu,\nu}},
\end{equation}
formula \eqref{eq:c-hat} takes the form
\begin{equation}
\hat{c}_{\mu,\nu} = \frac{1}{\hat{\rho}_{\mu,\nu}}\left(
\hat{h}_{\mu+1,\nu}\hat{\rho}_{\mu+1,\nu} + 
\hat{h}_{\mu,\nu}\hat{\rho}_{\mu-1,\nu} + 
\hat{h}_{\mu,\nu+1}\hat{\rho}_{\mu,\nu+1} + 
\hat{h}_{\mu,\nu}\hat{\rho}_{\mu,\nu-1}
\right),
\end{equation}
i.e., the function $\hat{\rho}_{\mu,\nu}$ is a solution of the equation
satisfied by $\hat\Phi_{m,n}$. This is the 5-point analogue of the fact that the
function $\tilde\theta_{m,n} = 1/\theta_{m,n}$ satisfies the
transformed equation \eqref{4p-affine}. Indeed, the function 
$\hat{\rho}_{\mu,\nu}$ is the restriction 
\begin{equation}
\hat\rho_{m,n} = \tilde\rho_{m-1,n}
\end{equation}
of the function
\begin{equation} 
\tilde\rho_{m,n} = 
\frac{\tilde\tau_{m,n}}{\tilde\tau_{m+1,n}}\tilde\theta_{m,n},
\end{equation}
obtained from $\tilde\theta_{m,n}$
in the same way as $\hat\Phi_{m,n}$ is obtained from 
$\tilde\psi_{m,n}$.
\end{Rem}
\subsection{DTs of the Schr\"{o}dinger 5-point scheme}
In what follows we will obtain the DT of the 5-point
Schr\"odinger scheme \cite{NSD} from the sub-lattice theory. In order not to
repeat the full procedure of the previous two subsections, we will use the
connection formulas \eqref{eq:con-ef-S} between the equal-field and
Schr\"{o}dinger gauges and the form \eqref{eq:def-Theta} 
of the particular solution
$\Theta_{\mu,\nu}$ of the Schr\"{o}dinger scheme. 
The transformed
potential $\hat\Gamma_{\mu,\nu}$ can be obtained from the connection
formulas \eqref{eq:con-ef-S} and the potential $\hat{h}_{\mu,\nu}$ of the
equal-field scheme given by equation \eqref{eq:def-h-h}, i.e.,
\begin{equation} \label{eq:DT-5p-Sch-Gamma}
\hat\Gamma_{\mu,\nu}^2 = \frac{1}{\hat{h}_{\mu,\nu}} =
\Gamma_{\mu,\nu} \Gamma_{\mu-1,\nu-1}
\frac{\Theta_{\mu-1,\nu-1}}{\Theta_{\mu,\nu}}.
\end{equation}
The transformed solution $\hat\Psi_{\mu,\nu}$, because of 
\eqref{eq:con-ef-S}, is related to $\hat\Phi_{\mu,\nu}$ by
\begin{equation}
\hat\Psi_{\mu,\nu} = \frac{\hat\Phi_{\mu,\nu}}{\hat\Gamma_{\mu,\nu}}.
\end{equation}
The transformed potential $\hat{F}_{\mu,\nu}$ 
can also be obtained from the connection
formulas \eqref{eq:con-ef-S} 
\begin{equation} \label{eq:DT-5p-Sch-hF}
\hat{F}_{\mu,\nu} =  \hat\Gamma_{\mu,\nu}^2 \hat{c}_{\mu,\nu} =
\frac{\Gamma_{\mu-1,\nu-1}\Theta_{\mu-1,\nu-1}}
{\Gamma_{\mu,\nu-1}\Theta_{\mu,\nu-1}} +
\frac{\Gamma_{\mu-1,\nu-1}\Theta_{\mu-1,\nu-1}}
{\Gamma_{\mu-1,\nu}\Theta_{\mu-1,\nu}} +
\frac{\Theta_{\mu,\nu}}{\Gamma_{\mu,\nu}}\left(
\frac{\Gamma_{\mu-1,\nu}}{\Theta_{\mu-1,\nu}} +
\frac{\Gamma_{\mu,\nu-1}}{\Theta_{\mu,\nu-1}}\right).
\end{equation}
Rewriting theorem \ref{th:DT-ef} in the new fields,
we obtain the DT of the Schr\"odinger 5-point scheme.
\begin{Th}
Let $\Psi_{\mu,\nu}$ and $\Theta_{\mu,\nu}$ be solutions of the
Schr\"odinger 5-point scheme \eqref{eq:Sch};
then the
solution $\hat\Psi_{\mu,\nu}$ of the following linear system of the first order
\begin{align} \label{eq:tSch1}
\frac{\hat\Gamma_{\mu+1,\nu} }{\Gamma_{\mu+1,\nu}}
\Theta_{\mu+1,\nu}\hat\Psi_{\mu+1,\nu} -
\frac{\hat\Gamma_{\mu,\nu} }{\Gamma_{\mu,\nu}}
\Theta_{\mu,\nu}\hat\Psi_{\mu,\nu}
& =
\;\;\frac{\Gamma_{\mu,\nu-1}}{\Gamma_{\mu,\nu}} \left(
\Theta_{\mu,\nu-1}\Psi_{\mu,\nu} - \Theta_{\mu,\nu}\Psi_{\mu,\nu-1}
\right) ,\\
\frac{\hat\Gamma_{\mu,\nu+1} }{\Gamma_{\mu,\nu+1}}
\Theta_{\mu,\nu+1}\hat\Psi_{\mu,\nu+1} -
\frac{\hat\Gamma_{\mu,\nu} }{\Gamma_{\mu,\nu}}
\Theta_{\mu,\nu}\hat\Psi_{\mu,\nu}
& =
-\frac{\Gamma_{\mu-1,\nu}}{\Gamma_{\mu,\nu}} \left(
\Theta_{\mu-1,\nu}\Psi_{\mu,\nu} - \Theta_{\mu,\nu}\Psi_{\mu-1,\nu},
\right)  \label{eq:tSch2},
\end{align}
where $\hat\Gamma_{\mu,\nu}$ is given by equation 
\eqref{eq:DT-5p-Sch-Gamma},
satisfies the 5-point Schr\"{o}dinger
scheme with the coefficients $\hat\Gamma_{\mu,\nu}$ and the coefficients
$\hat{F}_{\mu,\nu}$ given by \eqref{eq:DT-5p-Sch-hF}.
\end{Th}
\begin{Rem}
The analogue of the self-transformed solution $\hat\rho_{\mu,\nu}$, given by
equation \eqref{eq:def-h-rho}, of the
equal-field scheme is the solution
\begin{equation}
\hat\Theta_{\mu,\nu} = \frac{\hat\rho_{\mu,\nu}}{\hat\Gamma_{\mu,\nu}}  =
\frac{\Gamma_{\mu,\nu}}{\hat\Gamma_{\mu,\nu}\Theta_{\mu,\nu}}
\end{equation}
of the transformed Schr\"{o}dinger scheme.
In terms of $\hat\Theta_{\mu,\nu}$ the formula \eqref{eq:DT-5p-Sch-hF}
for the transformed potential
$\hat{F}_{\mu,\nu}$ takes the natural form
\begin{equation}\label{eq:DT-5p-Sch-f}
\hat{F}_{\mu,\nu} = \frac{1}{\hat\Theta_{\mu,\nu}}\left(
\frac{\hat\Gamma_{\mu,\nu}}{\hat\Gamma_{\mu+1,\nu}}\hat\Theta_{\mu+1,\nu}+
\frac{\hat\Gamma_{\mu-1,\nu}}{\hat\Gamma_{\mu,\nu}}\hat\Theta_{\mu-1,\nu}+
\frac{\hat\Gamma_{\mu,\nu}}{\hat\Gamma_{\mu,\nu+1}}\hat\Theta_{\mu,\nu+1}+
\frac{\hat\Gamma_{\mu,\nu-1}}{\hat\Gamma_{\mu,\nu}}\hat\Theta_{\mu,\nu-1}
\right).
\end{equation}
\end{Rem}

\section{Superposition of the DTs}
\label{sec:s-DTs}
Using the superposition principle
of the DTs of the discrete Moutard equation \cite{NimmoSchief}, 
we will obtain the corresponding
superposition principles for the DTs of the self-adjoint 5-point schemes in the
equal field gauge and in the Schr\"{o}dinger gauge.
\begin{Prop}
Given solutions $\psi_{m,n}$, $\theta^1_{m,n}$ and $\theta^2_{m,n}$ of the
4-point scheme \eqref{4p-affine}, denote by $\psi^{(1)}_{m,n}$ and
$\psi^{(2)}_{m,n}$ the transforms respectively of $\psi_{m,n}$ via
$\theta^1_{m,n}$ and $\theta^2_{m,n}$. If $\sigma_{m,n}$ is a
solution of the compatible linear system
\begin{align} \label{eq:4-DT-s1}
\sigma_{m+1,n} - \sigma_{m,n} & = \theta^2_{m,n}\theta^1_{m+1,n} -
\theta^1_{m,n}\theta^2_{m+1,n},\\ \label{eq:4-DT-s2}
\sigma_{m,n+1} - \sigma_{m,n} & = \theta^2_{m,n}\theta^1_{m,n+1} -
\theta^1_{m,n}\theta^2_{m,n+1},
\end{align}
then $\theta^{1(2)}_{m,n}$ and $\theta^{2(1)}_{m,n}$, given by
\begin{equation} \label{eq:sigma-t1-t2}
\theta^{1(2)}_{m,n}\theta^2_{m,n} = - \theta^{2(1)}_{m,n}\theta^1_{m,n}
= \sigma_{m,n},
\end{equation}
are solutions of the 4-points schemes satisfied respectively by
$\psi^{(2)}_{m,n}$ and $\psi^{(1)}_{m,n}$. Moreover, the
function $\psi^{(12)}_{m,n}$ given by
\begin{equation} \label{eq:4-DT-sup}
\psi^{(12)}_{m,n} - \psi_{m,n} =
\frac{\theta^1_{m,n}\theta^2_{m,n}}{\sigma_{m,n}}
(\psi^{(1)}_{m,n}-\psi^{(2)}_{m,n}),
\end{equation}
is simultaneously the transform of $\psi^{(1)}_{m,n}$ via
$\theta^{2(1)}_{m,n}$
and the transform of $\psi^{(2)}_{m,n}$ via $\theta^{1(2)}_{m,n}$. The
corresponding transformation of the $\tau$-function is given by
\begin{equation}
\tau^{(12)}_{m,n} = \sigma_{m,n}\tau_{m,n}.
\end{equation}
\end{Prop}
Our goal will be to derive the corresponding Bianchi superposition principle
for the 5-point equal-field scheme, and then for the Schr\"{o}dinger scheme. 
First, we will  rewrite equations \eqref{eq:4-DT-s1}-\eqref{eq:4-DT-s2} 
in terms of the transformation data
\begin{equation}
\rho^j_{m,n} = \frac{\tau_{m,n}}{\tau_{m+1,n}}\theta^j_{m,n}, \qquad j=1,2,
\end{equation}
of the equal-field gauge (see equation\eqref{eq:def-rho}),
and we will put them on the sub-lattice.

From equations \eqref{eq:4-DT-s1}-\eqref{eq:4-DT-s2} and the discrete
Moutard equation \eqref{4p-affine} we have 
\begin{align}
\sigma_{m,n-1} - \sigma_{m-1,n} & = - i
\frac{\tau_{m,n} \tau_{m-1,n-1}}{\tau_{m-1,n} \tau_{m,n-1}}\left(
\theta^1_{m,n}\theta^2_{m-1,n-1} - \theta^2_{m,n}\theta^1_{m-1,n-1}
\right), \\
\sigma_{m,n+1} - \sigma_{m-1,n} & = \; \; i 
\frac{\tau_{m,n} \tau_{m-1,n+1}}{\tau_{m-1,n} \tau_{m,n+1}}\left(
\theta^1_{m,n}\theta^2_{m-1,n+1} - \theta^2_{m,n}\theta^1_{m-1,n+1}
\right).
\end{align}
If we introduce the function
\begin{equation}
\Sigma_{m,n} =  i \sigma_{m-1,n}
\end{equation}
then the above equations can be rewriten in the form
\begin{align}
\Sigma_{m+1,n-1} - \Sigma_{m,n} & = \;\; 
h_{m,n} \left( \rho^1_{m,n} \rho^2_{m-1,n-1} - \rho^2_{m,n}\rho^1_{m-1,n-1}
\right),\\
\Sigma_{m+1,n+1} - \Sigma_{m,n} & = -
h_{m,n} \left( \rho^1_{m,n} \rho^2_{m-1,n+1} - \rho^2_{m,n}\rho^1_{m-1,n+1}
\right).
\end{align}

The functions $\Phi^{(j)}_{m,n}$, constructed according to equations 
\eqref{eq:def-h-Phi}-\eqref{eq:def-t-Phi}
\begin{equation}
\Phi^{(j)}_{m,n} = 
i\frac{\theta^j_{m-1,n}\psi^{(j)}_{m-1,n}}{\rho^j_{m,n}h_{m,n}},
\quad j=1,2,
\end{equation}
are the transforms of 
$\Phi_{\mu,\nu}$ via
$\rho^j_{\mu,\nu}$, when restricted to the sub-lattice. 
Consequently, the transformed data 
$\rho^{j(k)}_{\mu,\nu}$, $j\ne k$ are restrictions of the functions
\begin{equation}
\rho^{j(k)}_{m,n} = 
i\frac{\theta^k_{m-1,n}\theta^{j(k)}_{m-1,n}}{\rho^k_{m,n}h_{m,n}}, 
\quad j\ne k,
\end{equation}
which, due to equation \eqref{eq:sigma-t1-t2},
are related to $\Sigma_{m,n}$ by 
\begin{equation} \label{eq:Sigma-r1-r2}
h_{m,n}\rho^2_{m,n}\rho^{1(2)}_{m,n} = 
-h_{m,n}\rho^1_{m,n}\rho^{2(1)}_{m,n} =  \Sigma_{m,n}.
\end{equation}

Using equations
\eqref{eq:Mt1}-\eqref{eq:Mt2} we modify the superposition
formula \eqref{eq:4-DT-sup} in order to include
functions $\psi^{(1)}_{m-1,n}$ an $\psi^{(2)}_{m-1,n}$ instead of
$\psi^{(1)}_{m,n}$ and $\psi^{(2)}_{m,n}$
\begin{equation} \label{eq:4-DT-sup-mod}
\sigma_{m,n}\psi^{(12)}_{m,n} - \sigma_{m-1,n}\psi_{m,n} =
\theta^2_{m,n}\theta^1_{m-1,n}\psi^{(1)}_{m-1,n} -
\theta^1_{m,n}\theta^2_{m-1,n}\psi^{(2)}_{m-1,n}.
\end{equation}
Finally, applying twice equations \eqref{eq:def-h-Phi}-\eqref{eq:def-t-Phi}, 
we obtain that the function
\begin{equation}
\Phi^{(12)}_{m,n} = - \frac{\tau^{(12)}_{m-2,n}}{\tau^{(12)}_{m-1,n}}
\psi^{(12)}_{m-2,n},
\end{equation}
gives the desired superosition
$\Phi^{(12)}_{\mu,\nu}$, when restricted to the sub-lattice. 

Putting all the above considerations together, we obtain the
Bianchi-type permutability theorem, which can also be verified by direct
calculation.
\begin{Th} \label{th:DT-sup-ef}
Given solutions $\Phi_{\mu,\nu}$, $\rho^1_{\mu,\nu}$ and
$\rho^2_{\mu,\nu}$ of the
5-point equal field scheme \eqref{eq:5p-eq-f}, denote by
$\Phi^{(1)}_{\mu,\nu}$ and
$\Phi^{(2)}_{\mu,\nu}$ the transforms respectively of $\Phi_{\mu,\nu}$ via
$\rho^1_{\mu,\nu}$ and
$\rho^2_{\mu,\nu}$. If $\Sigma_{\mu,\nu}$ is a
solution of the compatible linear system
\begin{align} \label{eq:5-DT-e-f-1}
\Sigma_{\mu+1,\nu} - \Sigma_{\mu,\nu} & =
\; \; h_{\mu,\nu}
(\rho^1_{\mu,\nu} \rho^2_{\mu,\nu-1}-
\rho^2_{\mu,\nu} \rho^1_{\mu,\nu-1}) ,\\ \label{eq:5-DT-e-f-2}
\Sigma_{\mu,\nu+1} - \Sigma_{\mu,\nu} & =
- h_{\mu,\nu}
(\rho^1_{\mu,\nu} \rho^2_{\mu-1,\nu}-
\rho^2_{\mu,\nu} \rho^1_{\mu-1,\nu}) ,
\end{align}
then the functions $\rho^{1(2)}_{\mu,\nu}$ and $\rho^{2(1)}_{\mu,\nu}$ 
given by
\begin{equation}
h_{\mu,\nu}\rho^{1(2)}_{\mu,\nu}\rho^2_{\mu,\nu} =
- h_{\mu,\nu}\rho^{2(1)}_{\mu,\nu}\rho^1_{\mu,\nu}
= \Sigma_{\mu,\nu},
\end{equation}
are solutions of the 5-points schemes satisfied by
$\Phi^{(2)}_{\mu,\nu}$ and $\Phi^{(1)}_{\mu,\nu}$, correspondingly.
Moreover the function $\Phi^{(12)}_{\mu,\nu}$, given by
\begin{equation} \label{eq:DT-e-f-sup}
\frac{\Sigma_{\mu+1,\nu+1}}{\Sigma_{\mu,\nu}}\Phi^{(12)}_{\mu+1,\nu+1} +
\Phi_{\mu,\nu} =
\frac{h_{\mu,\nu}\rho^1_{\mu,\nu}\rho^2_{\mu,\nu}}{\Sigma_{\mu,\nu}}
(\Phi^{(2)}_{\mu,\nu} -
\Phi^{(1)}_{\mu,\nu}),
\end{equation}
is simultaneously the transform of $\Phi^{(1)}_{\mu,\nu}$ via
$\rho^{2(1)}_{\mu,\nu}$
and the transform of $\Phi^{(2)}_{\mu,\nu}$ via $\rho^{1(2)}_{\mu,\nu}$.
\end{Th}
\begin{Cor}
Equations \eqref{eq:def-h-h}
and \eqref{eq:c-hat}, together with the above theorem, imply
the transformation law of the coefficients of the doubly transformed equal
field scheme
\begin{equation}
h^{(12)}_{\mu+1,\nu+1}=
\frac{\Sigma_{\mu+1,\nu+1}}{ \Sigma_{\mu,\nu}}h_{\mu,\nu},
\end{equation}
\begin{multline}
c^{(12)}_{\mu+1,\nu+1} = 
\frac{\Sigma_{\mu+1,\nu+1}^2}{\Sigma_{\mu+1,\nu} \Sigma_{\mu,\nu+1}}
\Big( c_{\mu,\nu} + 
\frac{ h^2_{\mu,\nu}}{\Sigma_{\mu,\nu}}
\left( \rho^1_{\mu,\nu-1}\rho^2_{\mu-1,\nu} -
\rho^1_{\mu-1,\nu}\rho^2_{\mu,\nu-1} \right) +\\
+ \frac{h_{\mu+1,\nu} h_{\mu,\nu+1}}{\Sigma_{\mu+1,\nu+1}} 
\left( \rho^1_{\mu,\nu+1}\rho^2_{\mu+1,\nu} -
\rho^1_{\mu+1,\nu}\rho^2_{\mu,\nu+1} \right) 
\Big) .
\end{multline}
\end{Cor}

Finally, let us formulate the corresponding 
superposition principle for the
Schr\"{o}\-din\-ger scheme. It follows just from the connection 
formulas \eqref{eq:con-ef-S} and \eqref{eq:def-Theta} 
applied to theorem \ref{th:DT-sup-ef}.
\begin{Th}
Let $\Psi_{\mu,\nu}$, $\Theta^1_{\mu,\nu}$ and
$\Theta^2_{\mu,\nu}$ be solutions of the
5-point Schr\"{o}dinger scheme \eqref{eq:Sch}, denote by
$\Psi^{(1)}_{\mu,\nu}$ and
$\Psi^{(2)}_{\mu,\nu}$ the transforms respectively of $\Psi_{\mu,\nu}$ via
$\Theta^1_{\mu,\nu}$ and
$\Theta^2_{\mu,\nu}$, and let $\Gamma^{(1)}_{\mu,\nu}$,
$\Gamma^{(2)}_{\mu,\nu}$
denote the first coefficients of the corresponding equations, i.e.,
\begin{equation}
\left[\Gamma^{(j)}_{\mu,\nu}\right]^2 =
\frac{\Theta^j_{\mu-1,\nu-1}}{\Theta^j_{\mu,\nu}}
\Gamma_{\mu,\nu}\Gamma_{\mu-1,\nu-1}, \quad j=1,2.
\end{equation}
If $\Sigma_{\mu,\nu}$ is a
solution of the compatible linear system
\begin{align} \label{eq:5-DT-S1}
\Sigma_{\mu+1,\nu} - \Sigma_{\mu,\nu} & =\; \;
\frac{\Gamma_{\mu,\nu-1}}{\Gamma_{\mu,\nu}}
(\Theta^1_{\mu,\nu} \Theta^2_{\mu,\nu-1}-
\Theta^2_{\mu,\nu} \Theta^1_{\mu,\nu-1}) ,\\ \label{eq:5-DT-S2}
\Sigma_{\mu,\nu+1} - \Sigma_{\mu,\nu} & =
- \frac{\Gamma_{\mu-1,\nu}}{\Gamma_{\mu,\nu}}
(\Theta^1_{\mu,\nu} \Theta^2_{\mu-1,\nu}-
\Theta^2_{\mu,\nu} \Theta^1_{\mu-1,\nu}) ,
\end{align}
then $\Theta^{1(2)}_{\mu,\nu}$ and $\Theta^{2(1)}_{\mu,\nu}$, given by
\begin{equation}
\Theta^{1(2)}_{\mu,\nu}\Theta^2_{\mu,\nu}
\frac{\Gamma^{(2)}_{\mu,\nu}}{\Gamma_{\mu,\nu}} =
- \Theta^{2(1)}_{\mu,\nu}\Theta^1_{\mu,\nu}
\frac{\Gamma^{(1)}_{\mu,\nu}}{\Gamma_{\mu,\nu}}
= \Sigma_{\mu,\nu},
\end{equation}
are solutions of the 5-points schemes satisfied respectively by
$\Psi^{(2)}_{\mu,\nu}$ and $\Psi^{(1)}_{\mu,\nu}$.
The function $\Psi^{(12)}_{\mu,\nu}$, given by
\begin{equation} \label{eq:DT-Sch-sup}
\frac{\Sigma_{\mu+1,\nu+1}}{\Sigma_{\mu,\nu}}
\Gamma^{(12)}_{\mu+1,\nu+1}\Psi^{(12)}_{\mu+1,\nu+1} +
\Gamma_{\mu,\nu}\Psi_{\mu,\nu} =
\frac{ \Theta^1_{\mu,\nu}\Theta^2_{\mu,\nu}}{\Sigma_{\mu,\nu}}
(\Gamma^{(2)}_{\mu,\nu}\Psi^{(2)}_{\mu,\nu} -
\Gamma^{(1)}_{\mu,\nu}\Psi^{(1)}_{\mu,\nu}),
\end{equation}
where
\begin{equation} \label{eq:G-DT-s}
\left[\Gamma^{(12)}_{\mu+1,\nu+1}\right]^2=
\frac{\Sigma_{\mu,\nu}}{ \Sigma_{\mu+1,\nu+1}}\Gamma_{\mu,\nu}^2,
\end{equation}
is simultaneously the transform of $\Psi^{(1)}_{\mu,\nu}$ via
$\Theta^{2(1)}_{\mu,\nu}$
and the transform of $\Psi^{(2)}_{\mu,\nu}$ via $\Theta^{1(2)}_{\mu,\nu}$.
It satisfies the 5-point Schr\"{o}dinger scheme \eqref{eq:Sch} with
the coefficient $\Gamma^{(12)}_{\mu,\nu}$ given in \eqref{eq:G-DT-s}, 
and with the
coefficient $F^{(12)}_{\mu,\nu}$ given by
\begin{multline}
F^{(12)}_{\mu+1,\nu+1} = 
\frac{\Sigma_{\mu+1,\nu+1}\Sigma_{\mu,\nu}}
{\Sigma_{\mu+1,\nu}\Sigma_{\mu,\nu+1}}
\Big( F_{\mu,\nu} + 
\frac{ \Gamma_{\mu-1,\nu}\Gamma_{\mu,\nu-1}}
{\Sigma_{\mu,\nu}\Gamma_{\mu,\nu}^2}
\left( \Theta^1_{\mu,\nu-1}\Theta^2_{\mu-1,\nu} -
\Theta^1_{\mu-1,\nu}\Theta^2_{\mu,\nu-1} \right) +\\
+ \frac{\Gamma_{\mu,\nu}^2}{\Sigma_{\mu+1,\nu+1}
\Gamma_{\mu+1,\nu}\Gamma_{\mu,\nu+1}}  
\left( \Theta^1_{\mu,\nu+1}\Theta^2_{\mu+1,\nu} -
\Theta^1_{\mu+1,\nu}\Theta^2_{\mu,\nu+1} \right) 
\Big) .
\end{multline}
\end{Th}

\section{Algebro - geometric solutions}
\label{sec:al-geom}

In this sections we construct algebro - geometric solutions of the 4-point 
scheme
(\ref{4p-affine}) and, due to the results of the previous section, of the
5-point scheme (\ref{5p-rot}),(\ref{5p-affine}).

The direct and inverse periodic transform for a generic 4-point scheme
\eqref{eq:4p-gen} was developed in \cite{Kri2}. To obtain
algebro-geometric solutions of a specific 4-point scheme we have to impose 
suitable
constraints on the spectral data. As we shall see, the constraints which 
give rise
to the 4-point scheme (\ref{4p-affine}) are analogous to those introduced 
in
\cite{NV} in the study of the 2D continuous
Schr\"odinger operator. Constraints of such type were first introduced in 
\cite{Cher}
in the theory of reductions of 1+1 systems. In the theory
of discrete systems, analogous reductions were used in \cite{AKV} to 
characterize
orthogonal and Egoroff quadrilateral lattices.

We remark that algebro-geometric 5-point schemes were also studied in 
\cite{Kri2}.
The main difference between our 5-point schemes and the schemes studied in
\cite{Kri2} is the following. In our case, the full Bloch variety for zero 
energy
is constructed and the eigenfunctions with other energies are not 
incorporeted
in the spectral transform. Moreover, only self-adjoint schemes are 
constructed.
In \cite{Kri2}, generic schemes with the following property were studied:
for each eigenvalue, a finite number of eigenfunctions are explicitly 
constructed.

The standard finite-gap construction \cite{Kri2} for a generic 4-point 
scheme is based
on the following spectral data. Assume that we have:
\begin{enumerate}
\item A compact, regular, connected Riemann surface $\Gamma$ of genus $g$.
\item $l+1$ points $R_1$, \ldots, $R_{l+1}$ in $\Gamma$ -- the 
normalization points
for the wave function (denoted by $\Psi$).
\item $l+g$ points $\gamma_1$, \ldots, $\gamma_{l+g}$ in $\Gamma$ -- the 
divisor of
poles of the wave function.
\item A collection of points $P^+_1$, \ldots, $P^+_M$,  $P^-_1$, \ldots, 
$P^-_M$,
$Q^+_1$, \ldots, $Q^+_N$,  $Q^-_1$, \ldots, $Q^-_N$, where $M$, $N$ are 
arbitrary
positive integers.
\end{enumerate}
From the Riemann-Roch theorem it follows that, for generic data, there 
exists an
unique function $\Psi(\gamma,m,n)$, where $\gamma\in\Gamma$, 
$m,n\in{\mathbb Z}$ ,
$1\le m \le M$, $1\le n \le N$, with the following properties:
\begin{enumerate}
\item $\Psi(\gamma,m,n)$ is a meromorphic function of $\gamma$ in $\Gamma$.
\item $\Psi(\gamma,m,n)$ has at most first-order poles at the points 
$\gamma_k$,
$k=1,\ldots,g+l$,  $P^+_k$, $k=1,\ldots, m$, $Q^+_k$, $k=1,\ldots, n$ 
and no other
singularities.
\item $\Psi(\gamma,m,n)$ has at least first-order zeroes at the points
$P^-_k$, $k=1,\ldots, m$, $Q^-_k$, $k=1,\ldots, n$.
\item $\Psi(R_k,m,n)=1$, $k=1,\ldots, l+1$.
\end{enumerate}
The discrete time shift $m\rightarrow m+1$ corresponds to adding one 
extra pole
$P^+_{m+1}$ and one extra zero $P^-_{m+1}$ to the wave function.

Let us check that function $\Psi(\gamma,m,n)$ satisfies the 4-point equation
\begin{equation}
\label{eq:4p-wave}
\Psi(\gamma,m+1,n+1)+ \alpha_1(m,n) \Psi(\gamma,m+1,n) +
\alpha_2(m,n) \Psi(\gamma,m,n+1) + \alpha_3(m,n) \Psi(\gamma,m,n)=0,
\end{equation}
in which the $\gamma$-independent coefficients $\alpha_{1}(m,n)$, 
$\alpha_{2}(m,n)$,
$\alpha_3(m,n)$ are defined by the following formulas:
\begin{equation}
\label{eq:4p-wave1}
\alpha_1(m,n)=-\lim\limits_{\gamma\rightarrow P^+_{m+1}}
\frac{\Psi(\gamma,m+1,n+1)} {\Psi(\gamma,m+1,n)}
\end{equation}
\begin{equation}
\label{eq:4p-wave2}
\alpha_2(m,n)=-\lim\limits_{\gamma\rightarrow Q^+_{n+1}}
\frac{\Psi(\gamma,m+1,n+1)} {\Psi(\gamma,m,n+1)}
\end{equation}
\begin{equation}
\label{eq:4p-red1}
\alpha_3(m,n)=-1-\alpha_1(m,n)-\alpha_2(m,n).
\end{equation}
Indeed the left-hand side of \eqref{eq:4p-wave} has the following properties:
\begin{enumerate}
\item It is a meromorphic function of $\gamma$ in $\Gamma$.
\item It has at most first-order poles at the points $\gamma_k$,
$k=1,\ldots,g+l$,  $P^+_k$, $k=1,\ldots, m$, $Q^+_k$, $k=1,\ldots, n$
and no other singularities. (Conditions \eqref{eq:4p-wave1}, 
\eqref{eq:4p-wave2}
mean exactly that the poles at the points $P^+_{m+1}$, $Q^+_{n+1}$ vanish).
\item It has at least first-order zeroes at the points
$P^-_k$, $k=1,\ldots, m$, $Q^-_k$, $k=1,\ldots, n$ and $R_k$, 
$k=1,\ldots, l+1$.
\end{enumerate}
From the Riemann-Roch theorem it follows that, for generic data, this 
function
is identically equal to 0.

\begin{Rem}
In the paper \cite{Kri2} the points  $P^+_k$, $P^-_k$, $Q^+_k$, $Q^-_k$ 
were
generic. The special case in which all the $P^+_k$'s coincide: $P^+_k=P^+$
for $k=1,\ldots, M$, as well as all the $P^-_k$'s , $Q^+_k$'s, $Q^-_k$'s:
$P^-_k=P^-$, $k=1,\ldots, M$, $Q^+_k=Q^+$, $Q^-_k=Q^-$, $k=1,\ldots,N$
was also discussed in \cite{Kri2}. In this case, the discrete time
shift $m\rightarrow m+1$ corresponds to increasing the order of the pole
at the point $P^+$ and the order of the zero at the point $P^-$ of one,
and the Abel transform in the $\theta$-functional formulas (see below)
becomes constant. This last choice results in an essential effectivization
of the explicit formulas, but it corresponds to a more restricted class of 
potentials.
For example, if the potentials are periodic in $m$ with period $M$, the
Floquet multiplier at the point $Q^-_{n+1}$ can be calculated by the following
formula
$$
\kappa_1(n)=(-1)^M \prod\limits_{m=1}^M \frac{\alpha_3(m,n)}{\alpha_1(m,n)}.
$$
If all points $Q^-_{n}$ coincide, then $\kappa_1(n)$ should not depend on $n$.

Since we do not like to impose such a restriction on the class of our potentials,
we assume a generic collections of poles and zeroes.
\end{Rem}

\subsection{Discrete Moutard reductions and constraints on the spectral data}
Let us describe the reductions corresponding to the affine 4-point scheme
\eqref{4p-affine} (may be with a complex potential $f(m,n)$).

\begin{Lem}
\label{4p-lemma1}
Assume that $\Gamma$
possess a holomorphic involution $\sigma$ with exactly 2 fixed points $R_+=R_1$,
$R_-$. Assume that the spectral data have the following symmetry with respect to
$\sigma$:
\begin{enumerate}
\item Exactly one normalization point $R_1=R_+$ is used ($l=0$).
\item There exists a meromorphic differential $\Omega$ with 2 first-order poles
at the fixed points $R_+$, $R_-$ and $2g$ zeroes at the points $\gamma_1$, \ldots,
$\gamma_g$, $\sigma\gamma_1$, \ldots, $\sigma\gamma_g$.
\item $\sigma P^+_k=P^-_k$, $\sigma Q^+_k=Q^-_k$ for all $k$.
\end{enumerate}
Then
\begin{equation}
\label{eq:4p-red2}
\alpha_3(m,n)+ 1 = 0, \ \ \alpha_1(m,n) + \alpha_2(m,n)= 0
\end{equation}
and, consequently, $\Psi(\gamma,m,n)$ satisfies the 4-point scheme \eqref{4p-affine}
with
\begin{equation}
\label{eq:4p-red2.5}
f(m,n)=i\alpha_1(m,n)=-i\!\!\!\!\lim\limits_{\gamma\rightarrow P^+_{m+1}}
\!\!\!\frac{\Psi(\gamma,m+1,n+1)} {\Psi(\gamma,m+1,n)}=
i\!\!\!\!\lim\limits_{\gamma\rightarrow Q^+_{n+1}}
\!\!\!\frac{\Psi(\gamma,m+1,n+1)} {\Psi(\gamma,m,n+1)}.
\end{equation}
\end{Lem}

\begin{proof} By analogy with \cite{NV}, consider the
following form:
$$
\Omega \Psi(\gamma,m,n) \Psi(\sigma\gamma,m,n).
$$
This form has 2 first-order poles at the points $R_+$, $R_-$ and no other
singularities. Therefore
$$
\res(\Omega \Psi(\gamma,m,n) \Psi(\sigma\gamma,m,n),\gamma=R_+)=
-\res( \Omega \Psi(\gamma,m,n) \Psi(\sigma\gamma,m,n),{\gamma=R_-})
$$
Taking into account that
$$
\res(\Omega ,\gamma=R_+)=
-\res( \Omega ,{\gamma=R_-}),
$$
we obtain
$$
\Psi^2(R_-,m,n)=\Psi^2(R_+,m,n)\equiv1.
$$
Let us show that
\begin{equation}
\label{eq:4p-red3}
\Psi(R_-,m,n)=(-1)^{m+n}.
\end{equation}
If $m=n=0$ we know that $\Psi(\gamma,0,0)\equiv 1$ (from the uniqueness argument
and taking into account that the constant 1 satisfies all the analyticity constraints).
Therefore $\Psi(R_-,0,0)=1$.
Consider the function $\Psi(R_-,m+1,n)$ as
a function of the point $P^+_{m+1}$ (we use here that $P^-_{m+1}=\sigma P^+_{m+1}$).
By construction this function is meromorphic in $P^+_{m+1}$. Since the surface
$\Gamma$ is connected, then $\Psi(R_-,m+1,n)$ does not depend on $P^+_{m+1}$.
Assume that $P^+_{m+1}$ be located very close to $R_-$ and denote by $z$ a local
coordinate in a neighbourhood of $R_-$, $z(R_-)=0$, such that $\sigma z = -z$. Then
$$
\Psi(z,m+1,n)=\frac{z+P^+_{m+1}}{z-P^+_{m+1}}\Psi(z,m,n)+ o(P^+_{m+1});
$$
therefore
$$
\Psi(R_-,m+1,n)=-\Psi(R_-,m,n).
$$

Substituting \eqref{eq:4p-red3} into \eqref{eq:4p-wave} and taking into account
\eqref{eq:4p-red1}, we obtain \eqref{eq:4p-red2}. This completes the proof.
\end{proof}

Having in mind applications in discrete geometry and in discrete complex functions
theory, let us formulate sufficient conditions which guaranty that the potential
$f(m,n)$ be real.
\begin{Lem}
\label{4p-lemma2}
Assume that all the constraints of Lemma~\ref{4p-lemma1} on the spectral 
data be
fulfilled. Let the spectral data satisfy the following 
additional restrictions.
There exists an antiholomorphic involution $\tau$ on $\Gamma$ with the following
properties:
\begin{enumerate}
\item $\tau$ commutes with $\sigma$.
\item $\tau R_+=R_-$.
\item The points  $P^+_k$, $P^-_k$, $Q^+_k$,  $Q^-_k$ are fixed point of $\tau$.
\item The set of divisor points $\gamma_k$ is invariant under $\tau$ (but $\tau$
may map one divisor point into another).
\end{enumerate}
Then the potential $f(m,n)$ and the function $\Psi(\gamma,m,n)$ have the
following reality properties:
\begin{equation}
\label{eq:4p-red4}
\bar f(m,n)=f(m,n), \ \
\Psi(\tau\gamma,m,n)= (-1)^{m+n} \overline{ \Psi(\gamma,m,n)}.
\end{equation}
In particular, if $\gamma$ is a
real point of $\Gamma$: $\tau\gamma=\gamma$, then $\Psi(\gamma,m,n)$ is either
real or pure imaginary:
\begin{equation}
\label{eq:4p-red5}
\begin{array}{l}
\Psi(\gamma,m,n)\in {\mathbb R} \ \ \ \mbox{if} \ \ m+n \ \ \mbox{is even} \\
\Psi(\gamma,m,n)\in i{\mathbb R} \ \ \mbox{if} \ \  m+n \ \ \mbox{is odd}.
\end{array}
\end{equation}
\end{Lem}
To prove the Lemma, it is sufficient to point out that the function $(-1)^{m+n}
\overline{\Psi(\tau\gamma,m,n)} $ satisfies all the analytic constraints
characterizing $\Psi(\gamma,m,n)$. Therefore these two functions coincide.
The reality of $f(m,n)$ then follows immediately from \eqref{eq:4p-red2.5}.

\begin{Rem} The reality constraint on the eigenfunction $\Psi(\gamma,m,n)$
naturally distinguishes between even and odd sub-lattices.
\end{Rem}

In Section 2 it was shown on the algebraic level that any solution of the 4-point scheme
\eqref{4p-affine} generates a solution of the 5-point scheme \eqref{5p-affine}.
In this section we give an analytic proof of this statement.

Consider the following meromorphic 1-form
$$
\tilde\Omega(\gamma,m,n,\tilde m,\tilde n)=\Omega(\gamma) \Psi(\sigma\gamma,m,n)
\Psi(\gamma,\tilde m,\tilde n).
$$
Assume that the differential $\Omega$ used before to define the constraints
on the spectral data has the following normalization:
$$
\res(\Omega,\gamma=R_+)=\frac12, \ \ \res(\Omega,\gamma=R_-)=-\frac12.
$$
Let us calculate the residues at the poles of $\tilde\Omega(\gamma,m,n,\tilde m,\tilde
n)$ for $|\tilde m - m|\le 1$, $|\tilde n - n|\le 1$. It is convenient to write
the answer in graphic form. We denote our Riemann surface by an oval,
and we assume that our distinguished points are located at the following positions:
\begin{center}
\mbox{\hsize5cm \vbox{\epsfxsize=5cm
\epsffile{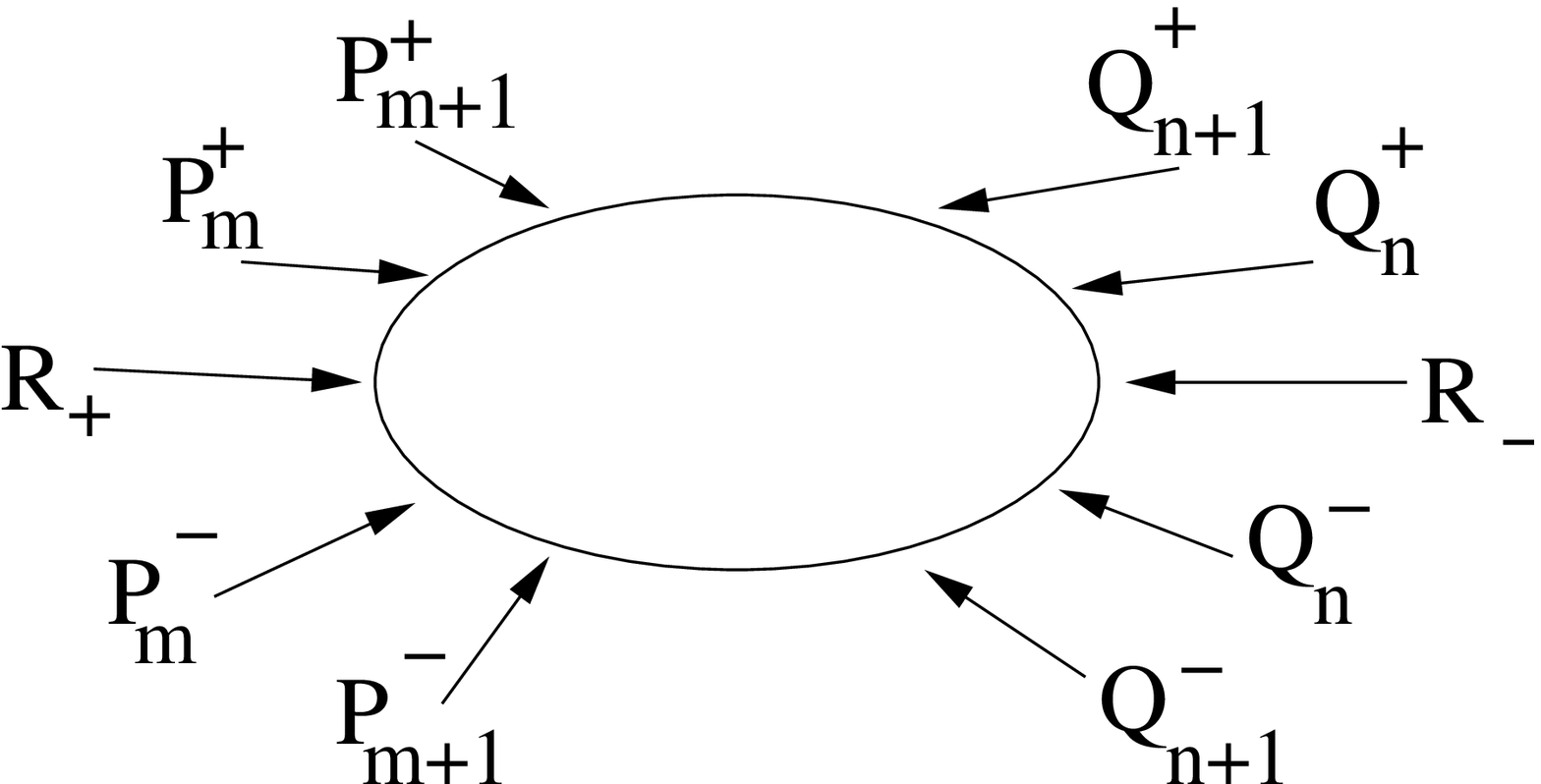}}}
\end{center}

In the next picture we show the poles of $\tilde\Omega(\gamma,m,n,\tilde m,\tilde
n)$. We use the following notations: $+$ denotes a pole with residue $+\frac12$,
$-$ denotes a pole with residue $-\frac12$, $\times$ denotes a pole with a residue
different from $\pm\frac12$, $\circ$ denotes a zero. The parameter $\tilde n$ is
written in the left column, the parameter $\tilde m$ is written in the bottom row.
The residues are written next to the symbols $\times$.

\begin{center}
\mbox{\hsize12cm \vbox{\epsfxsize=12cm
\epsffile{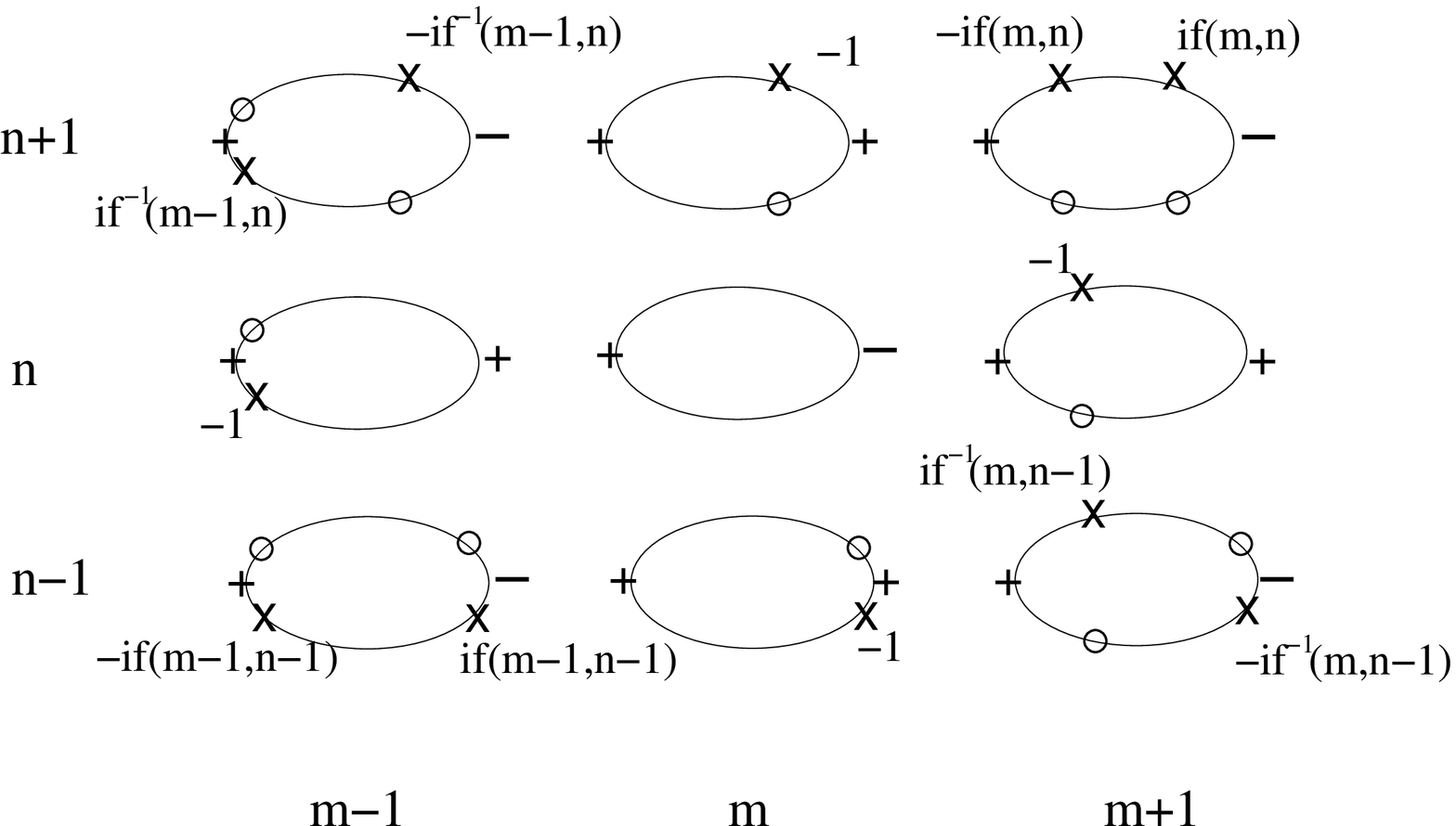}}}
\end{center}

Let us explain how the residues are calculated. First of all, for $\tilde m=m$,
$\tilde n=n$ we have only 2 poles and the residues are opposite, therefore we have
an identity (which was used when we proved the Lemma~~\ref{4p-lemma1}). If $\tilde m=m\pm1$,
$\tilde n=n$ or $\tilde m=m$,$\tilde n=n\pm 1$ we have 3 poles. At $R_+$ and $R_-$
the residues are $+\frac12$, therefore the 3rd residue is $-1$. If
$|\tilde m - m|=|\tilde n - n|=1$, the residue at $R_+$ is equal to $+\frac12$,
the residue at $R_-$ is equal to $-\frac12$, therefore we have 2 more poles
with opposite residues. For example, let us calculate the residues for
$\tilde m=m+1$,$\tilde n=n+1$. From the 4-point scheme \eqref{4p-affine}
it follows that
$$
\res(\tilde\Omega(\gamma,m,n,m+1,n+1),\gamma=P^+_{m+1})=
if(m,n)\res(\tilde\Omega(\gamma,m,n,m+1,n),\gamma=P^+_{m+1}),
$$
therefore
\begin{equation}
\label{4p-res1}
\res(\tilde\Omega(\gamma,m,n,m+1,n+1),\gamma=P^+_{m+1})=-if(m,n).
\end{equation}
All the other residues can be calculated exactly in the same way.

To prove that the function $\Psi(\gamma,m,n)$ satisfies the 5-point scheme
\begin{equation}
\label{5p-bloch}
\begin{array}{l}
\frac{1}{f(m,n})(\Psi(\gamma,m+1,n+1)-\Psi(\gamma,m,n))+\\
\frac{1}{f(m-1,n-1)}(\Psi(\gamma,m-1,n-1)-\Psi(\gamma,m,n))+\\
f(m,n-1)(\Psi(\gamma,m+1,n-1)-\Psi(\gamma,m,n))+\\
f(m-1,n)(\Psi(\gamma,m-1,n+1)-
\Psi(\gamma,m,n))=0
\end{array}
\end{equation}
it is enough to verify that the left-hand side of \eqref{5p-bloch} has the
poles and zeroes prescribed for $\Psi(\gamma,m,n)$ and, moreover it is equal
to zero at the points $R_+$, $R_-$. Therefore, by the Riemann-Roch theorem,
it is zero.

\subsection{Algebro-geometrical solutions}

Riemann surfaces with the constraints described above can be constructed in the
following way. We start from the ``vacuum'' solution, corresponding to the Riemann
sphere $g=0$:
$$
if(m,n)=\frac{P_{m+1}+Q_{n+1}}{P_{m+1}-Q_{n+1}}, \ \
\Psi(\lambda,m,n) = \prod\limits_{k=1}^m\frac{\lambda+P_{k}}{\lambda-P_{k}}
\prod\limits_{k=1}^n\frac{\lambda+Q_{k}}{\lambda-Q_{k}}.
$$

If all $P_k$'s coincide as well as all $Q_k$'s: $P_k=P$, $Q_k=Q$, the above vacuum solution
reduces to the so-called ``discrete exponential''
\begin{equation}
\label{d-exp}
e(m,n;\lambda) =\left(\frac{\lambda+P}{\lambda-P}\right)^m\left(\frac{\lambda+Q}{\lambda-Q}\right)^n
\end{equation}
discussed in \cite{BobMerSu}; if $Q=iP$, then $f(m,n)\equiv 1$ and (\ref{d-exp}) solves the
Cauchy-Riemann equation (\ref{d-CR}).

The involutions $\sigma$ and $\tau$ are given by:
$$
\sigma\lambda = -\lambda, \ \ \tau\lambda=\frac{1}{\bar\lambda}
$$
and the Riemann sphere with all its distinguished points is drawn here:

\begin{center}
\mbox{\hsize5cm \vbox{\epsfxsize=5cm
\epsffile{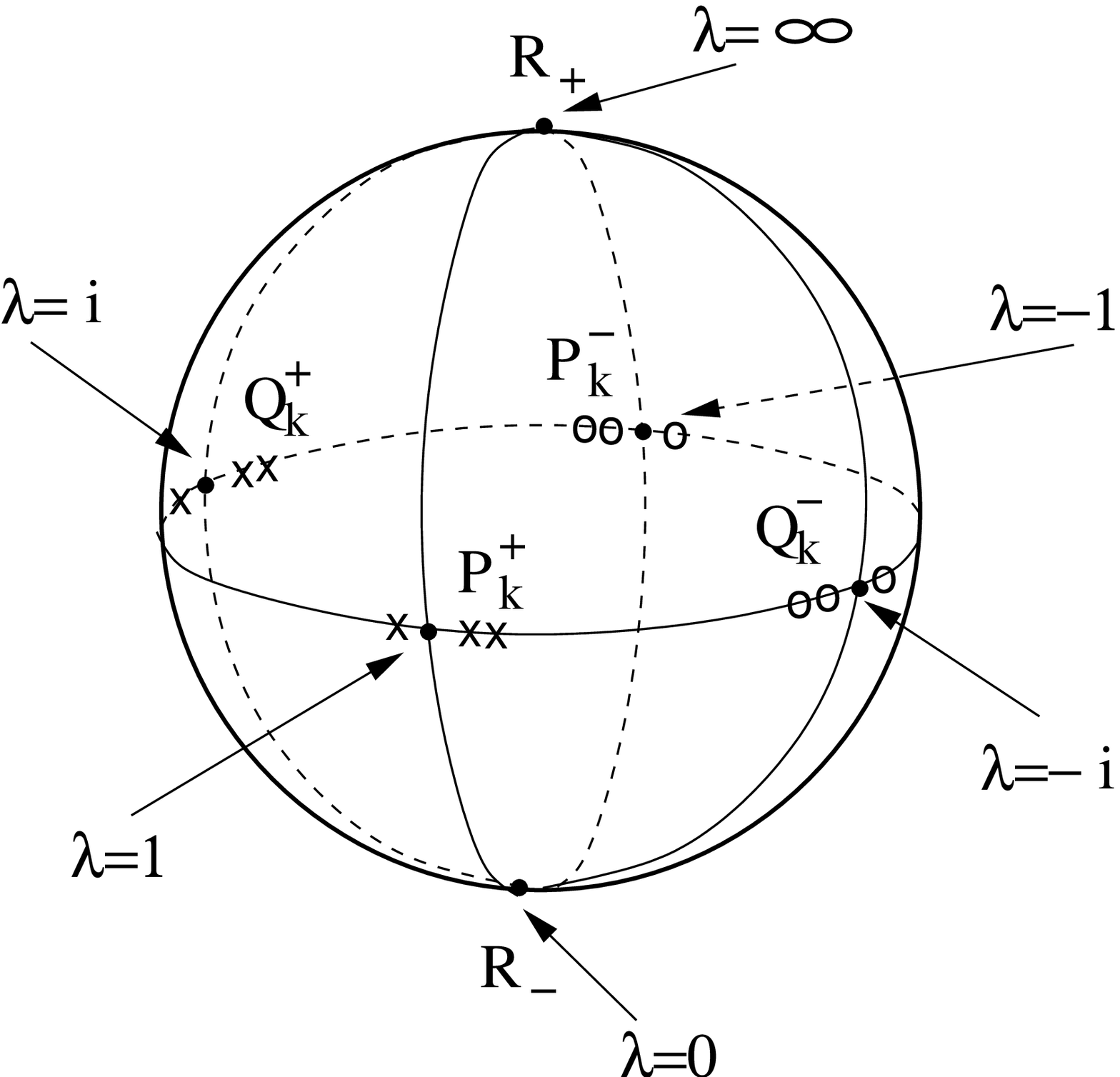}}}
\end{center}

By analogy with the continuous case, discrete holomorphic polynomials can be constructed from
the discrete exponential (\ref{d-exp}) in the following way (if $Q=iP$):
\begin{equation}
\label{d-poly1}
p^{(j)}(m,n)=\frac{1}{(2P)^j}\left.\frac{d^je(m,n;1/\zeta)}{d\zeta^j}\right|_{\zeta=0}.
\end{equation}

The first 3 examples read:
\begin{equation}
\label{d-poly2}
p^{(0)}=1, \ \ p^{(1)}=m+in, \ \  p^{(2)}=(m+in)^2, \ \  p^{(3)}=(m+in)^3+(m-in)/2.
\end{equation}
We observe that the discrete holomorphic polynomials coincide with the continuous monomials $z^j$
(assuming $z=m+in$) up to the order 2.

A non-trivial Riemann surface can be constructed by attaching handles to the above
Riemann sphere keeping the above symmetries. Two examples are shown below.
\begin{center}
\mbox{\hsize5cm \vbox{\epsfxsize=5cm
\epsffile{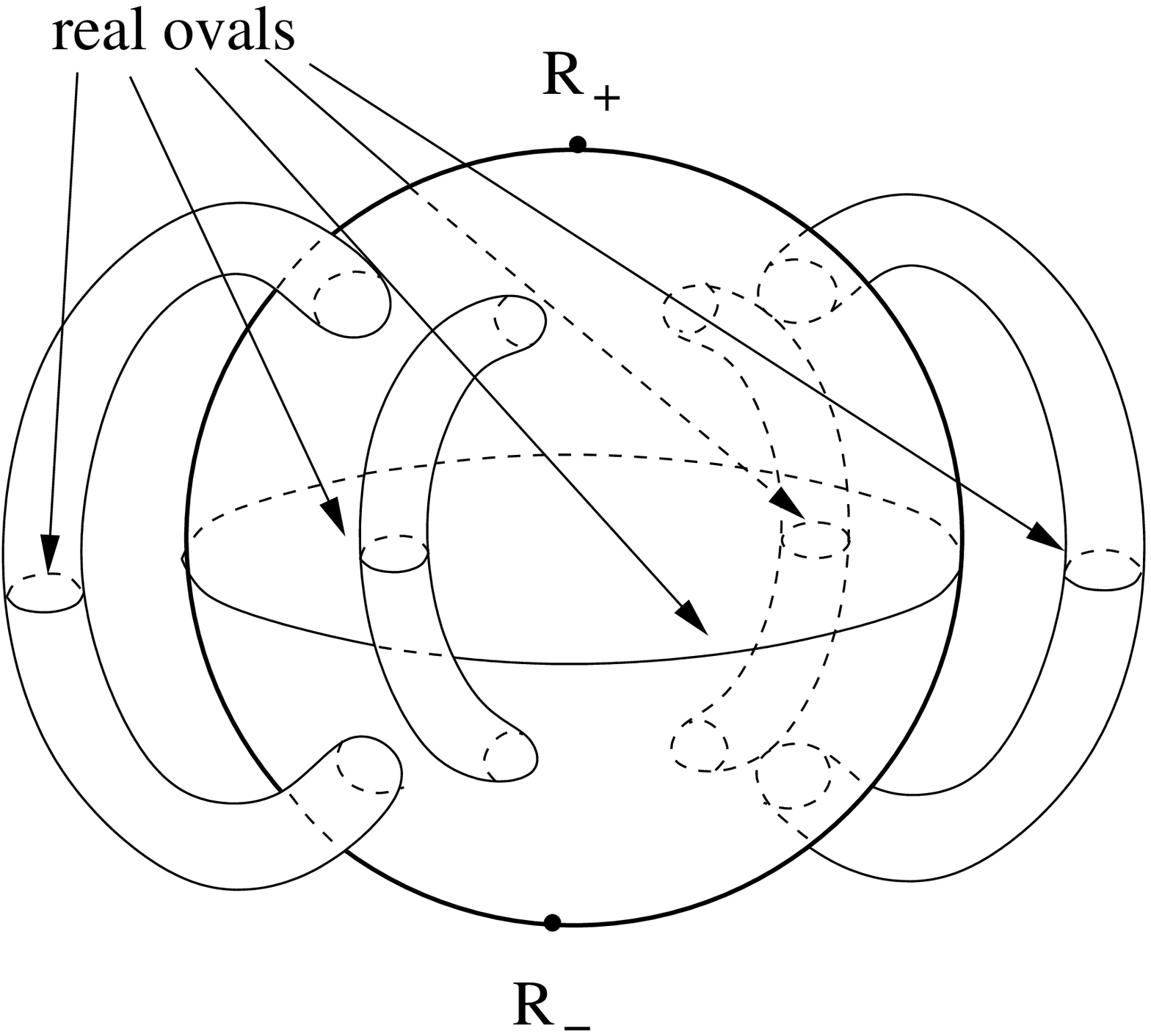}}}
\mbox{\hsize5cm \vbox{\epsfxsize=5cm
\epsffile{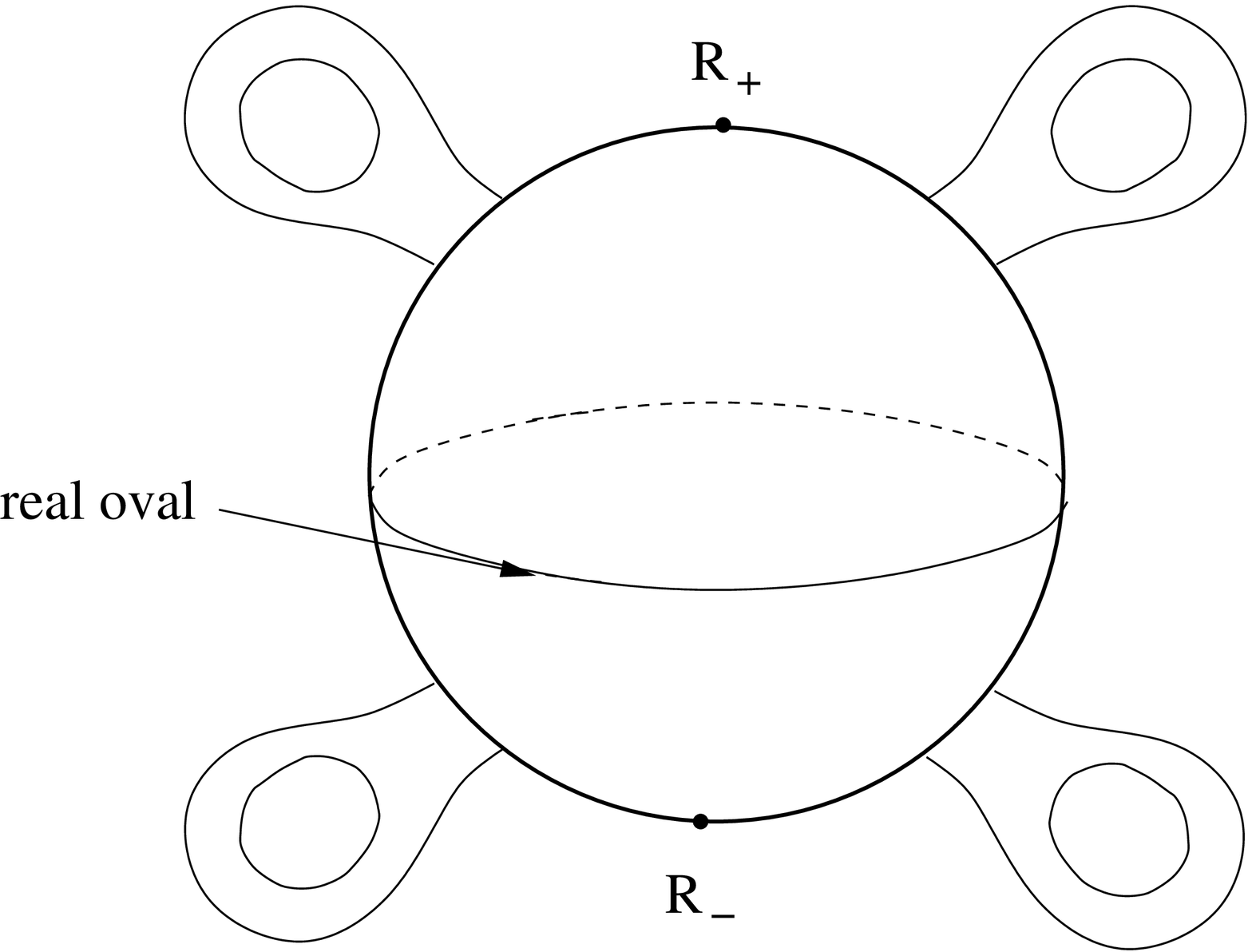}}}
\end{center}
The figure on the left illustrates the so-called $M$-curve (the number of real
ovals, i.e. the ovals formed by the fixed points of the antiholomorphic
involution $\tau$, is equal $g+1$. $g+1$ is the greatest possible number of real
ovals. In this particular case $g=4$).  The figure on the right
illustrates a curve with $g=4$ and only one real oval.

\begin{Rem}
In the theory of the 2-D continuous Schr\"odinger operator the $M$-curves play
a distinguished role. In particular the operators generated by the $M$-curves are
non-singular and strictly positive \cite{NV}, \cite{GN}. A complete classification
of finite-gap data corresponding to nonsingular solutions was obtained in
\cite{Natan}. The study of analogous properties of the 4-point and 5-point schemes
will be the subject of future investigation.
\end{Rem}

It is well-known that function $\Psi(\gamma,m,n)$ can be written in terms
of Riemann $\theta$-functions in the following way:

$$
\Psi(\gamma,m,n)=\ \ \ \ \ \ \ \ \ \ \ \ \ \ \ \ \ \ \ \ \ \ \ \
\ \ \ \ \ \ \ \ \ \ \ \ \ \ \ \ \ \ \ \ \ \ \ \ \ \ \ \ \ \ \ \ \ \
\ \ \ \ \ \ \ \ \ \ \ \ \ \ \ \ \ \ \ \ \ \ \ \ \ \ \ \ \ \ \
$$
\begin{equation}
\label{Psi}
=\frac{\theta\left(\left.\vec A(\gamma)\! + \! \sum\limits_{k=1}^m
(\vec A(P^-_k) \! - \! \vec A(P^+_k)) \! + \!
\sum\limits_{k=1}^n( \vec A(Q^-_k) \! - \! \vec A(Q^+_k)) \!
- \sum\limits_{k=1}^g \! \vec A(\gamma_k) \! - \! \vec K\right|B \right)}
{\theta\left(\left.\vec A(\gamma)\! - \! \sum\limits_{k=1}^g \! \vec A(\gamma_k) \! - \!
\vec K\right|B \right)}\times
\end{equation}
$$
\times\frac{\theta\left(\left.\vec A(R_+) - \! \sum\limits_{k=1}^g \! \vec A(\gamma_k) \! - \!
\vec K\right|B\right)}
{\theta\left(\left.\vec A(R_+)+\sum\limits_{k=1}^m
(\vec A(P^-_k) \! - \! \vec A(P^+_k)) \! + \!
\sum\limits_{k=1}^n( \vec A(Q^-_k) \! - \! \vec A(Q^+_k)) \!
- \sum\limits_{k=1}^g \! \vec A(\gamma_k) \! - \! \vec K\right|B\right)} \times
$$
$$
\times\exp\left( \sum\limits_{k=1}^m\int\limits_{R_+}^{\gamma}
\Omega(\tilde\gamma,P^+_k,P^-_k)
+\sum\limits_{k=1}^n\int\limits_{R_+}^{\gamma}\Omega(\tilde\gamma,Q^+_k,Q^-_k) \right).
$$
Here we have used the following notations. $a_k$, $b_k$ are the basic cycles in
$\Gamma$, $a_k\circ b_l=\delta_{kl}$, $a_k\circ a_l=b_k\circ b_l=0$. $\omega_i$
are the basic holomorphic differentials such that

\begin{equation}
\oint_{a_j} \omega_i = \delta_{ij},
\label{fg-n1}
\end{equation}
$\Omega(\gamma,P,Q)$ are meromorphic differentials of the third kind with 2
first-order poles in $P$, $Q$, with residues $-1$ and $+1$ respectively and
zero $a$-periods:
\begin{equation}
\res\bigl(\Omega(\gamma,P,Q),\gamma=P \bigr)= -1, \ \
\res\bigl(\Omega(\gamma,P,Q),\gamma=Q \bigr)= 1,
\oint_{a_j} \Omega(\gamma,P,Q) = 0.
\end{equation}
$\vec A(\gamma)=(A_1(\gamma),..,A_g(\gamma))$ denotes the Abel transform:
\begin{equation}
A_k(\gamma)= \int\limits_{{\cal P}}^\gamma \omega_k, \ \
k=1, \ldots , g,
\end{equation}
where the starting point ${\cal P}$ of the Abel transform can be chosen
arbitrarely. The Riemann $\theta$-functions are defined by the following Fourier series:
\begin{equation}
\label{theta}
\theta(\vec z|B) = \sum_{m_1, \ldots, m_g} \exp \left \{ \pi i
\sum_{kj} B_{kj}m_k m_j + 2 \pi i\sum_k z_k m_k \right \},
\end{equation}
where $B_{kl}$ are the $b$-periods of the holomorphic differentials:
\begin{equation}
\oint_{b_j} \omega_i =B_{ji}
\label{fg-n5}
\end{equation}
(in the definition of the $\theta$-function we use the same normalizations  as in \cite{Mumford};
other classical books, like \cite{Fay}, use the different normalization corresponding to a-periods
equal to $2\pi i$).

Due to the symmetry of the spectral curve imposed by $\sigma$, the Riemann
$\theta$-functions can be expressed as bilinear combinations of Riemann and Prym
$\theta$-functions of genus $g/2$ \cite{Fay}. This is analogous to the 
continuous Schr\"odinger
case \cite{NV}.

The simplest non-trivial example corresponds to an $M$-curve with 2 handles.
\begin{center}
\mbox{\hsize5cm \vbox{\epsfxsize=5cm
\epsffile{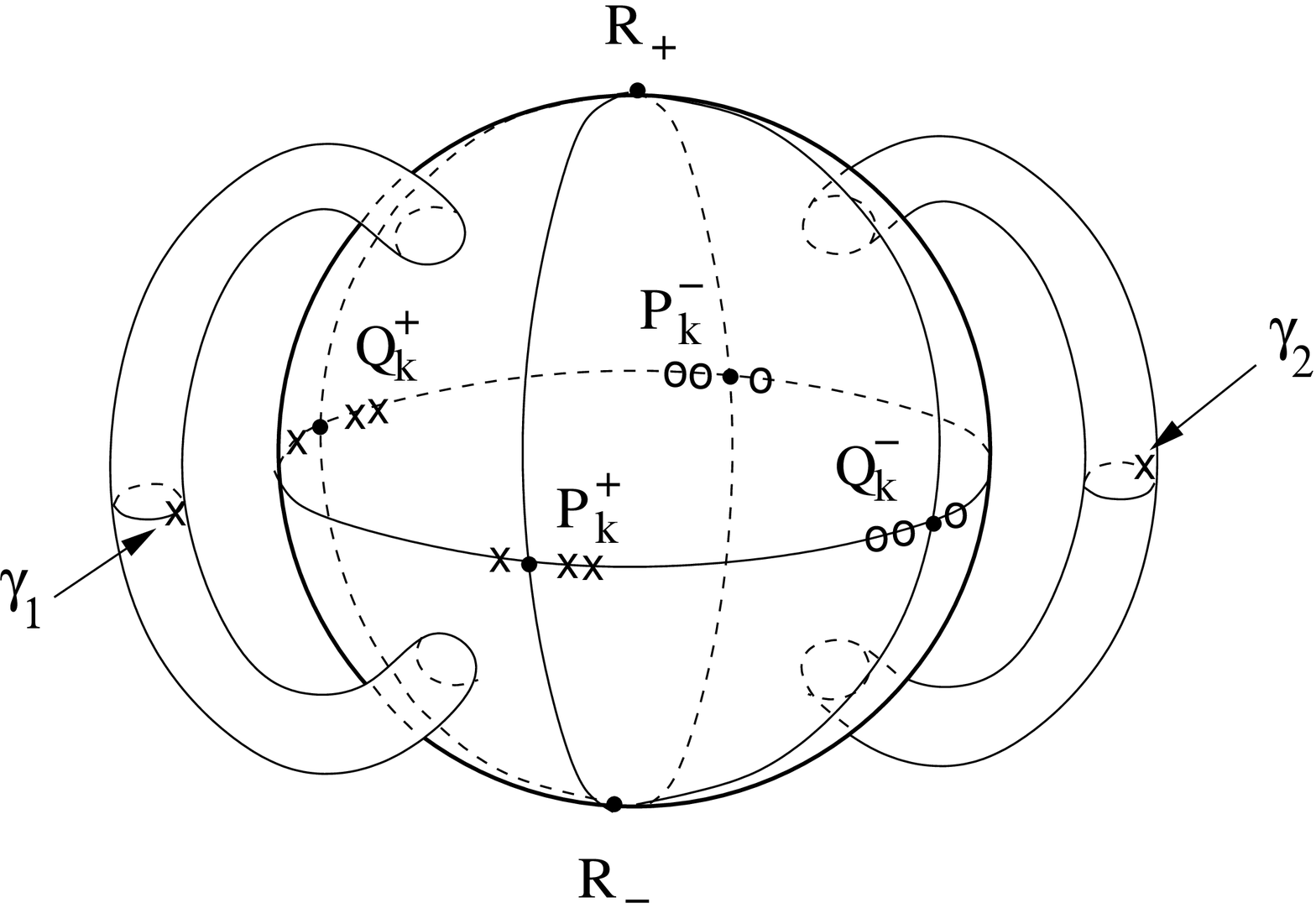}}}
\end{center}
In this case, due to the symmetry imposed by $\sigma$, the answer can be expressed in terms of genus 1
$\theta$-functions. It is convenient to choose a basis of cycles with the following
symmetry (see \cite{Fay}):
\begin{equation}
\sigma a_1= - a_2, \ \ \sigma b_1=-b_2.
\end{equation}
Denote by $\omega$ and $\hat\omega$ the holomorphic differentials such that
\begin{equation}
\label{Symm1}
\sigma\omega=\omega, \ \ \sigma\hat\omega=-\hat\omega, \ \ \oint\limits_{a_1} \omega=
\oint\limits_{a_1} \hat\omega=1
\end{equation}
and let
\begin{equation}
\eta= \oint\limits_{b_1} \omega, \ \ \hat\eta= \oint\limits_{b_1} \hat\omega.
\end{equation}
Then
\begin{equation}
\omega_1=\frac{\hat\omega+\omega }{2},\ \ \omega_2=\frac{\hat\omega-\omega }{2},
B_{11}=B_{22}=\frac{\hat\eta+\eta }{2}, \ \ B_{12}=B_{21}=\frac{\hat\eta-\eta }{2},
\end{equation}
and
\begin{equation}
\label{theta2}
\begin{array}{c}
\theta(z_1,z_2|B)=\theta(z_1+z_2|2\eta)\theta(z_1-z_2|2\hat\eta)+ \\
\theta(z_1+z_2+\eta|2\eta)\theta(z_1-z_2+\hat\eta|2\hat\eta)
\exp{\left[\pi i \left(\frac{\eta+\hat\eta}{2}+2z_1 \right)\right]}.
\end{array}
\end{equation}

Let us consider the following explicit example in which the Riemann surface $\Gamma$ is defined by the
equation:
\begin{equation}
\label{hyperelliptic}
\mu^2=\prod\limits_{k=1}^6 (\lambda-e_k), \ \ \ \ e_{7-k}=-e_k,
\end{equation}
$R_+$, $R_-$ are the infinite points $\mu\sim\lambda^3$ and  $\mu\sim -\lambda^3$ respectively,
with $\lambda\sim\infty$,
$\sigma:(\lambda,\mu)\rightarrow(-\lambda,-\mu)$,
$\tau:(\lambda,\mu)\rightarrow(\bar\lambda,-\bar\mu)$, $\omega$ and $\hat\omega$ are defined by:
\begin{equation}
\omega=d_1\frac{d\lambda}{i\mu}, \ \ \hat\omega=d_2\frac{\lambda d\lambda}{i\mu},
\end{equation}
where the parameters $d_1$, $d_2$ are fixed by the normalization condition (\ref{Symm1}). The
$a$ and $b$ cycles are shown in the figure below, the cuts are along the intervals $[e_1,e_2]$,
$[e_3,e_4]$,  $[e_5,e_6]$, the solid lines correspond to the sheet containing $R_+$,
the dashed lines correspond to the sheet containing $R_-$.

\begin{center}
\mbox{\hsize10cm \vbox{\epsfxsize=10cm
\epsffile{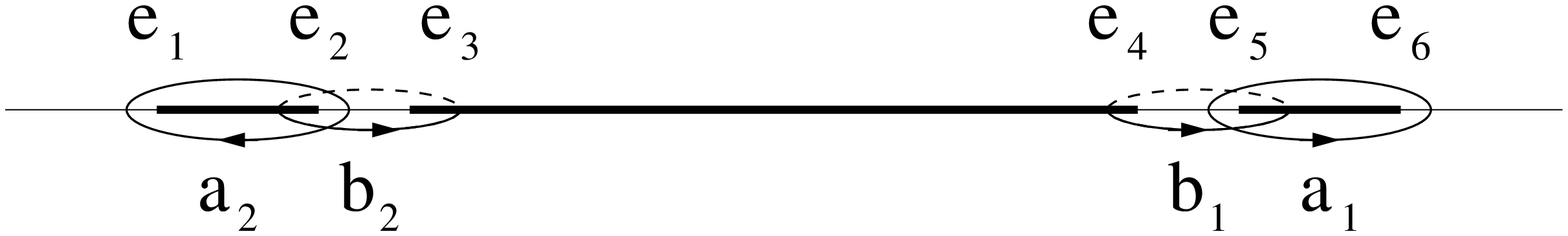}}}
\end{center}

The divisor is located on the ovals $a_1$, $a_2$, the distinguished points $P_k^+$,
$P_k^-$, $Q_k^+$, $Q_k^-$, are located on the oval lying over the interval  $[e_3,e_4]$.

If the starting point of the Abel transform is chosen at one of the branch points of a
hyperelliptic Riemann surface, the vector of Riemann constants can be written in a very
simple form (see \cite{Fay}). If ${\cal P}=e_6$, then
\begin{equation}
\vec K=\left[\begin{array}{c} \frac12+\frac{B_{12}}{2} \\ \frac12+\frac{B_{22}}{2} \end{array} \right].
\end{equation}

\subsection{Some applications}

As it was already mentioned in the introduction, the 4-point and 5-point schemes (\ref{4p-affine}) and
(\ref{5p-affine}) are relevant in the theory of discrete complex functions, as well as in
the theory of integrable discrete geometries.

Solutions $\Psi(\gamma,m,n)$ of equation (\ref{4p-affine}) provide
examples of discrete holomorphic functions and, through the 
change of variables (\ref{eq:mn-mn}),
generate discrete harmonic functions on the sub-lattice. Let us concentrate here
our attention, for instance, on the discrete analogues of a polynomial of degree $2$.

If $g=2$, the corresponding discrete holomorphic function is
constructed from the $\Psi(\gamma,m,n)$ function (\ref{Psi}),(\ref{theta}) associated
with the hyperelliptic curve (\ref{hyperelliptic}), via the formula:
\begin{equation}
\label{DiscH2}
\lim_{\gamma\to R_+}\left(\left[\Psi(\gamma,m,n)+\Psi(\sigma\gamma,m,n)
-2\Psi(R_+,m,n)\right]\frac{\lambda^2}{4}\right).
\end{equation}

Due to the positions of the distinguished points $P^+$, $Q^+$, the  corresponding $g=0$ discrete holomorphic function
has the following form:
\begin{equation}
\label{DiscH0}
\Re\left(e^{i\frac43\pi} p^{(2)}(m,n) \right)=  \Re\left(e^{i\frac43\pi} (m+in)^2 \right).
\end{equation}

The graphs of the real parts of the discrete holomorphic functions (\ref{DiscH0}) and (\ref{DiscH2})
are shown in the figure below.

\begin{center}
\parbox{7cm}{\hsize7cm \vbox{\epsfxsize=7cm \epsffile{Holom0.eps}} ~\\ \begin{center} $g=0$ \end{center} }
\parbox{7cm}{\hsize7cm \vbox{\epsfxsize=7cm \epsffile{Holom1.eps}} ~\\
\begin{center} $g=2$ \end{center}  }
\end{center}
\noindent For $g=2$ the spectral data are: $e_1=-3$,  $e_2=-2.02$, $e_3=-2$,
$\gamma_1=(2.6,i\sqrt{16.56635904} ) $, $\gamma_2=(-2.6,i\sqrt{16.56635904})$,
$P^+=(-1,i\sqrt{73.9296})$,  $Q^+=(-\sqrt{3},-i\sqrt{6.4824})$.
\medskip

As it was shown in \cite{NS}, solutions $\Psi(\gamma,\mu,\nu)$ 
of equation (\ref{5p-affine})
allow to construct quadrilateral surfaces in $\RR^3$, i.e., discrete surfaces whose
elementary quadrilaterals are planar, in the following way.

Consider the vector $\vec N_{\mu,\nu}:~\ZZ^2\to\RR^3$ whose components are three independent real solutions
of the self-adjoint 5-point scheme (\ref{5p-affine}). Then $\vec N_{\mu,\nu}$ is the normal
vector of a quadrilater surface $\vec r_{\mu,\nu}:~\ZZ^2\to~\RR^3$ defined by the
following generalization of the Lelieuvre formulas:
\begin{equation}
\label{Lel}
\Delta_\mu \vec r_{\mu,\nu} = -b_{\mu,\nu-1} \N_{\mu,\nu} \times \N_{\mu,\nu-1},~~~~
\Delta_\nu \vec r_{\mu,\nu} = a_{\mu-1,\nu} \N_{\mu,\nu} \times \N_{\mu-1,\nu}.
\end{equation}

Here we consider examples of quadrilateral surfaces constructed using the discrete analogues
of complex polynomials of degree 1. If $g=0$, we consider the following normal vector:
\begin{equation}
{\vec N}^{(0)}_{\mu,\nu}=\left[2\Re\left(e^{i\frac23\pi} (\mu-\nu+i(\mu+\nu)) \right) ,
2\Im\left(e^{i\frac23\pi} (\mu-\nu+i(\mu+\nu)) \right),1 \right].
\end{equation}
Analogously, if $g=2$, the normal vector is defined by
\begin{equation}
{\vec N}^{(2)}_{\mu,\nu}=\left[\Re\left(\lim_{\gamma\to R_+}\left(\left[\Psi(\gamma,\mu-\nu,\mu+\nu)-
1\right]\lambda \right)\right),
\Im\left(\lim_{\gamma\to R_+}\left(\left[\Psi(\gamma,\mu-\nu,\mu+\nu)- 1\right]\lambda \right)\right),1 \right],
\end{equation}
where $\Psi$ is the wave function (\ref{Psi}),(\ref{theta}) associated with the hyperelliptic curve
(\ref{hyperelliptic}).
Then the two quadrilateral lattices constructed, via the embedding (\ref{Lel}), by the
normal vectors ${\vec N}^{(0)}_{\mu,\nu}$ and $\tilde{\vec N}^{(2)}_{\mu,\nu}$, are shown in the
figures below.
\begin{center}
\parbox{7cm}{\hsize7cm \vbox{\epsfxsize=7cm \epsffile{Surf0.eps}} ~\\ 
\begin{center} $g=0$ \end{center} }
\parbox{7cm}{\hsize7cm \vbox{\epsfxsize=7cm \epsffile{Surf1.eps}} ~\\
\begin{center} $g=2$ \end{center}  }
\end{center}


\section*{Acknowledgments}

\noindent
This work was supported by three cultural and
scientific agreements: that between the University of Roma ``La Sapienza'' 
and the University
of Warsaw, that between the University of Roma ``La Sapienza'' and the 
Landau Institute
of Theoretical Physics, and that between the University of Roma 
``La Sapienza'' and the
University of Warmia and Mazury in Olsztyn. A. D. and M. N. were partially
supported by the KBN grant 2~P03B~126~22. P.G. was also supported by the
Russian Foundation for Basic Research grant No 02-01-00659 and by the
Russian Science Support Foundation.

\end{document}